\documentclass{article}

\usepackage[cmex10]{amsmath}
\usepackage{amssymb}
\usepackage[all]{xypic}
\usepackage{latexsym}
\usepackage{amsfonts} 
\usepackage{enumerate} 

\newtheorem{theorem}{\bfseries Theorem}[section]
\newtheorem{lemma}[theorem]{\bfseries Lemma}
\newtheorem{corollary}[theorem]{\bfseries Corollary}
\newtheorem{defn}[theorem]{\bfseries Definition}
\newtheorem{prop}[theorem]{\bfseries Proposition}

\newcommand{\ti}[1]{\textit{#1}}
\newcommand{\tbf}[1]{\textbf{#1}}
\newcommand{\mc}[1]{\mathcal{#1}}

\newcommand{\cl}[1]{\ensuremath{\mathcal{#1}}}

\newcommand{\Blacksquare}{\rule{.7em}{.7em}}
\newcommand{\epf}{\hfill $\Blacksquare$}

\newcommand{\lt}{{{\L}o{\'s}-Tarski}}

\newcommand{\PS}{\ensuremath{\mathbb{PS}}}
\newcommand{\PSC}{\ensuremath{\mathbb{PSC}}}

\newcommand{\trees}{Trees}
\newcommand{\words}{Words}
\newcommand{\myr}{r}
\newcommand{\s}{s}
\newcommand{\myt}{t}

\newcommand{\f}{\ensuremath{\mathsf{f}}}
\newcommand{\g}{\ensuremath{\mathsf{g}}}
\newcommand{\mf}[1]{\ensuremath{\mathfrak{#1}}}
\newcommand{\eat}[1]{}
\newcommand{\Children}{\ensuremath{\mathsf{Children}}}
\begin{document}

\title{A Generalization of the {\lt} Preservation Theorem over Classes
  of Finite Structures}

\author{Abhisekh Sankaran, Bharat Adsul, Supratik Chakraborty\vspace{3pt}\\
\small{Department of Computer Science and Engineering},\vspace{-3pt}\\
\small{IIT Bombay, Mumbai, India}}

\maketitle
\thispagestyle{empty}

\begin{abstract}
We investigate a generalization of the {\lt} preservation theorem via
the semantic notion of \emph{preservation under substructures modulo
  $k$-sized cores}. It was shown earlier that over arbitrary
structures, this semantic notion for first-order logic corresponds to
definability by $\exists^k\forall^*$ sentences.  In this paper, we
identify two properties of classes of finite structures that ensure
the above correspondence. The first is based on well-quasi-ordering
under the embedding relation.  The second is a logic-based
combinatorial property that strictly generalizes the first.  We show
that starting with classes satisfying any of these properties, the
classes obtained by applying operations like disjoint union, cartesian
and tensor products, or by forming words and trees over the classes,
inherit the same property. As a fallout, we obtain interesting classes
of structures over which an effective version of the {\lt} theorem
holds.
\end{abstract}


\section{Introduction}\label{section:intro}

Preservation theorems in first-order logic (henceforth called FO) have
been extensively studied in model theory \cite{chang-keisler}. A FO
preservation theorem asserts that the collection of FO definable
classes closed under a model-theoretic operation corresponds to the
collection of classes definable by a syntactic fragment of FO.  A
classical preservation theorem is the {\lt} theorem, which states that
over \emph{arbitrary structures}, a FO sentence is preserved under
substructures iff it is equivalent to a universal
sentence~\cite{chang-keisler}.  In~\cite{wollic12-paper}, it was
conjectured that the {\lt} theorem can be generalized using a simple
yet delicate semantic notion of a class of structures being
\emph{preserved under substructures modulo $k$-sized cores}.  This
semantic notion, denoted $\PSC(k)$ and explained in detail in
Section~\ref{section:prelims}, is parameterized by a quantitative
model-theoretic parameter $k$.  For $k=0$, this reduces to the usual
notion of preservation under substructures.  The conjecture
in~\cite{wollic12-paper} was settled in \cite{arxiv-2013}, where it
was shown that over arbitrary structures, a FO sentence is preserved
under substructures modulo $k$-sized cores iff it is equivalent to a
$\exists^k\forall^*$ sentence. This result, which we abbreviate as
$PSC(k)=\exists^k\forall^*$, provides a non-trivial generalization of
the {\lt } theorem.

Since classes of finite structures are the most interesting from a
computational point of view, researchers have studied preservation
theorems over finite structures in the
past~\cite{dawar-pres-under-ext, duris-ext, dawar-hom,
  rossman-hom}.  Most preservation theorems, including the {\lt}
theorem, fail over the class of all finite structures\footnote{A
  notable exception is the homomorphism preservation theorem
  \cite{rossman-hom}.}.  Recent works
\cite{dawar-pres-under-ext,duris-ext} have studied structural and
algorithmic properties of classes of finite structures that allow the
{\lt } theorem to hold over these classes.  Unfortunately, these
studies don't suffice to identify classes over which
$PSC(k)=\exists^k\forall^*$ holds.  In this paper, we try to fill this
gap by formulating and studying new abstract properties of classes of
finite structures.
As a fallout of our studies, we not only identify interesting classes
over which $PSC(k) = \exists^k\forall^*$ holds for all $k \ge 0$, but
also identify classes that lie beyond those studied
by~\cite{dawar-pres-under-ext, duris-ext} and yet satisfy the {\lt}
theorem.

The remainder of the paper is organized as follows.  In
Section~\ref{section:prelims}, we discuss preliminaries and set up the
notation.  Section~\ref{section:pwqo-and-plogic} introduces a
well-quasi-ordering based property and a logic-based combinatorial
property of classes of finite structures, where both properties are
parameterized by a natural number $k$.  We show that
$PSC(k)=\exists^k\forall^*$ holds over classes satisfying these
properties. We also formulate an `effective' version of the
logic-based property that allows us to compute from a sentence defining a class
in $PSC(k)$, an equivalent $\exists^k\forall^*$ sentence.  
In Section~\ref{section:relations}, we undertake an
exhaustive comparison of the collections of classes that the
aforementioned properties define.  We show in
Section~\ref{section:words-and-trees} that the classically interesting
and well-studied classes of words and trees over a finite alphabet
belong to these collections.  Finally, in
Section~\ref{section:constructing-new-classes}, we establish
composition theorems for the above collections of classes under a set
of natural composition operators such as disjoint union, cartesian and
tensor products. We further show that the above collections are also
closed under the operations of forming words and trees. The results in
Sections \ref{section:words-and-trees} and
\ref{section:constructing-new-classes} are amongst the most
technically involved results in this paper. Throughout, we provide
examples of interesting classes satisfying the various properties
discussed.

\section{Notation and Preliminaries}\label{section:prelims}
Let $\mathbb{N}$ denote the natural numbers \emph{including zero}.  We
assume that the reader is familiar with standard notation and
terminology of first-order logic.  We consider only finite
vocabularies, represented by $\tau$, that are relational (i.e. contain
only predicate and constant symbols).  Standard notions of
\emph{$\tau$-structures, substructures} and \emph{extensions} (see
\cite{chang-keisler}) are used throughout.  \emph{All
  $\tau$-structures considered in this paper are assumed to be
  finite.}  Given a $\tau$-structure $\mathfrak{A}$, we use
$\mathsf{U}_\mathfrak{A}$ to denote the universe of $\mathfrak{A}$ and
$|\mf{A}|$ to denote its cardinality or size.  If $A$ is a subset of
$\mathsf{U}_{\mathfrak{A}}$, we use $\mathfrak{A}(A)$ to denote the
substructure of $\mathfrak{A}$ induced by $A$.  Given
$\tau$-structures $\mathfrak{A}$ and $\mathfrak{B}$, we use
$\mathfrak{A} \subseteq \mathfrak{B}$ to denote that $\mf{A}$ is a
substructure of $\mf{B}$.  If $A$ and $B$ are sets, we also use $A
\subseteq B$ to denote set containment.  We say that $\mathfrak{A}$
\emph{embeds} in $\mathfrak{B}$ if $\mathfrak{A}$ is isomorphic to a
substructure of $\mathfrak{B}$.  Notationally, we represent this as
$\mathfrak{A} \hookrightarrow \mathfrak{B}$.  It is easy to see that
$\hookrightarrow$ is a pre-order over any class of $\tau$-structures.
\emph{All classes of $\tau$-structures, and subclasses thereof,
  considered in this paper are assumed to be closed under
  isomorphism.}

We denote by $FO(\tau)$ the set of all FO formulae over $\tau$.  A
sequence $(x_1, \ldots, x_k)$ of variables is written as $\bar{x}$.
For notational convenience, we abbreviate a block of quantifiers of
the form $Q x_1 \ldots Q x_k$ by $Q^k\bar{x}$, where $Q \in \{\forall,
\exists\}$.  Given a $\tau$-structure $\mathfrak{A}$ and a $FO(\tau)$
sentence $\varphi$, if $\mathfrak{A} \models \varphi$, we say that
$\mf{A}$ is a \emph{model} of $\varphi$.  Given a class $\cl{S}$ of
$\tau$-structures of interest, every $FO(\tau)$ sentence $\varphi$
defines a unique subclass of $\cl{S}$ consisting of all models of
$\varphi$.  Therefore, when $\cl{S}$ is clear from the context, we
interchangeably talk of a set of $FO(\tau)$ sentences and the
corresponding collection of subclasses of $\cl{S}$.

The notion of a class of $\tau$-structures being preserved under
substructures modulo bounded cores was introduced
in~\cite{wollic12-paper}.  This notion is central to our work.  The
following is an adapted version of the definition given
in~\cite{wollic12-paper}.  
\begin{defn}\label{defn:PSC(k)}
Let $\cl{S}$ be a class of $\tau$-structures and $k \in \mathbb{N}$.
A subclass $\cl{C}$ of $\cl{S}$ is said to be \emph{preserved under
  substructures modulo $k$-sized cores over $\cl{S}$} if every
$\tau$-structure $\mathfrak{A} \in \cl{C}$ has a subset
$\mathsf{Core}$ of $\mathsf{U}_{\mathfrak{A}}$ such that (i)
$|\mathsf{Core}| \le k$, and (ii) for every $\mf{B} \in \cl{S}$, if
$\mathfrak{B} \subseteq \mathfrak{A}$ and $\mathsf{Core} \subseteq
\mathsf{U}_{\mathfrak{B}}$, then $\mathfrak{B} \in \cl{C}$.  The set
$\mathsf{Core}$ is called a \emph{$k$-core of $\mathfrak{A}$ with
  respect to $\cl{C}$ over $\cl{S}$}.
\end{defn}
As an example, if $\cl{S}$ represents the class of all graphs, the
subclass $\cl{C}$ of acyclic graphs is preserved under substructures
module $k$-sized cores over $\cl{S}$, for every $k \ge 0$.  Like
Definition~\ref{defn:PSC(k)}, most other definitions, discussions and
results in this paper are stated with respect to an underlying class
$\cl{S}$ of structures.  For notational convenience, when $\cl{S}$ is
clear from the context, we omit mentioning ``over $\cl{S}$''.  In
Definition~\ref{defn:PSC(k)}, if the subclass $\cl{C}$ of $\cl{S}$ and
$k \in \mathbb{N}$ are also clear from the context, we call
$\mathsf{Core}$ simply as a core of $\mathfrak{A}$.

Given a class $\cl{S}$, let ${\PSC}(k)$ denote the collection of all
subclasses of $\cl{S}$ that are preserved under substructures modulo
$k$-sized cores.  As shown by the example of acyclic graphs above,
${\PSC}(k)$ may contain subclasses that are not definable over
$\cl{S}$ by any FO sentence.  Since our focus in this paper is on
classes definable by FO sentences, we define $PSC(k)$ to be the
collection of classes in ${\PSC}(k)$ that are definable over $\cl{S}$
by FO sentences.  As before, we interchangeably talk of $PSC(k)$ as a
collection of classes and as a set of the defining FO sentences.
Since ${\PSC}(0)$ coincides with the property of preservation under
substructures, we abbreviate ${\PSC}(0)$ as ${\PS}$ and $PSC(0)$ as
$PS$ in the remainder of the paper.

Given $k, p \in \mathbb{N}$, let $\exists^k\forall^p$ denote the set
of all $FO(\tau)$ sentences in prenex normal form whose quantifier
prefix has $k$ existential quantifiers followed by $p$ universal
quantifiers.  We use $\exists^k\forall^*$ to denote $\bigcup_{p \in
  \mathbb{N}} \exists^k\forall^p$.  As before, when the class $\cl{S}$
of $\tau$-structures is clear from the context, we use
$\exists^k\forall^p$ and $\exists^k\forall^*$ to also denote the
corresponding subclasses of $\cl{S}$.  We refer the reader to
\cite{wollic12-paper} for interesting examples from the collections
$\PS, PS, \forall^*, \PSC(k), PSC(k), \exists^k\forall^*$ and for
inclusion relationships among these collections.

Using the above notation, the {\lt} theorem can be stated as follows.
\begin{theorem}\label{theorem:los-tarski-primal}
Over arbitrary structures, $PS = \forall^*$.
\end{theorem}
In~\cite{arxiv-2013}, this was generalized to give the
following result.
\begin{theorem}\label{theorem:PSC(k)=E^kA*}
Over arbitrary structures, for every $k \in \mathbb{N}$, $PSC(k) =
\exists^k\forall^*$.
\end{theorem}
It is easy to see that if $\varphi$ is an $\exists^k\forall^*$
sentence and $\mf{A} \models \varphi$, then every witness of the
existential variables of $\varphi$ forms a $k$-core of $\mathfrak{A}$.
However, the converse is not necessarily true~\cite{wollic12-paper}.
Specifically, let $\tau = \{E\}$, where $E$ is a binary predicate.
Consider the $FO(\tau)$ sentence $\varphi ~\equiv~ \exists x\, \forall
y\; E(x, y)$, and the $\tau$-structure $\mathfrak{A}$ defined by
$\mathsf{U}_{\mathfrak{A}} = \{0, 1\}$ and $E^{\mathfrak{A}} = \{(0,
0), (0, 1), (1, 1)\}$.  Clearly, $\mathfrak{A} \models \varphi$ and
there is only one witness of the existential quantifier, viz. $0$.
However, both $\{0\}$ and $\{1\}$ are cores of $\mathfrak{A}$!

\vspace{2pt} In~\cite{wollic12-paper}, the notion of
\emph{relativizing} a FO sentence with respect to a finite set of
variables was introduced. We recall this for later use.  Let
$\mathsf{Const}$ be the set of constants in a relational vocabulary
$\tau$.  Given a sentence $\phi$ over $\tau$ and a sequence of
variables $\bar{x}$, let $\phi|_{\bar{x}}$ denote the quantifier-free
formula with free variables $\bar{x}$, obtained as follows.  Suppose
$X$ is the underlying set of $\bar{x}$.  We first replace every
$\forall$ in $\phi$ by $\neg \exists$, and then replace every
subformula of the form $\exists x\, \psi(x, y_1, \ldots, y_k)$ by
$\bigvee_{z \;\in\; X \,\cup\, \mathsf{Const}} \psi(z, y_1, \ldots,
y_k)$.  The formula $\phi|_{\bar{x}}$ is called \emph{$\phi$
  relativized to $X$}.  Informally, given a $\tau$-structure
$\mathfrak{A}$, the formula $\phi|_{\bar{x}}$ asserts that $\phi$ is
true in the substructure of $\mathfrak{A}$ induced by the underlying
set of $\bar{x}$. More precisely, for every $(a_1, \ldots a_k) \in
\mathsf{U}_{\mathfrak{A}}^k$, we have $(\mathfrak{A}, a_1, \ldots,
a_k) \models \phi|_{\bar{x}}$ iff $\mathfrak{A}(\{a_1, \ldots, a_k\})
\models \phi$.

As mentioned in Section~\ref{section:intro}, recent studies have
identified structural and algorithmic properties of classes of finite
structures that allow $PS = \forall^*$ to hold over these
classes~\cite{dawar-pres-under-ext,duris-ext}.  For example, the
class of structures whose Gaifman graph is acyclic was shown to admit
$PS = \forall^*$ in~\cite{dawar-pres-under-ext}.  Let $\cl{S}$ be the
class of graphs, where each graph is a disjoint union of finite
undirected paths.  Clearly, every such graph has an acyclic Gaifman
graph, and hence $PS = \forall^*$ over $\cl{S}$.  However, as shown
in~\cite{wollic12-paper}, $PSC(k) \neq \exists^k\forall^*$, for every
$k \ge 2$, over $\cl{S}$.  This motivates us to ask: \emph{Can we
  identify abstract properties of classes of finite structures that
  allow $PSC(k) = \exists^k\forall^*$ to hold over these classes?}
Our primary contribution is the identification of two properties that
answer the above question affirmatively.

\section{Two Properties of Classes of Structures}\label{section:pwqo-and-plogic}
We define two properties of classes of finite structures, each of
which entails $PSC(k) = \exists^k \forall^*$ over the class.

\subsection{A property based on well-quasi-orders}
Recall that a pre-order $(\Pi, \preceq)$ is \emph{well-quasi-ordered}
(w.q.o.)  if for every infinite sequence $\pi_1,\pi_2,\ldots$ of
elements of $\Pi$, there exists $i < j$ such that $\pi_i \preceq
\pi_j$ (see \cite{diestel}).  If $(\Pi, \preceq)$ is a w.q.o., we say that ``$\Pi$ is a
w.q.o. under $\preceq$''.  It is a basic fact that if $\Pi$ is a
w.q.o. under $\preceq$, then for every infinite sequence
$\pi_1,\pi_2,\ldots$ of elements of $\Pi$ there exists an
infinite subsequence $\pi_{i_1},\pi_{i_2},\ldots$ such that
$i_1 < i_2 < \ldots$ and $\pi_{i_1} \preceq \pi_{i_2} \preceq \ldots$.

Given a vocabulary $\tau$ and $k\in \mathbb{N}$, let $\tau_k$ denote
the vocabulary obtained by adding $k$ new constant symbols to $\tau$.
Let $\cl{S}$ be a class of structures.  We use $\cl{S}_k$ to denote the
class of all $\tau_k$-structures whose $\tau$-reducts are structures
in $\cl{S}$.  Our first property, denoted $\mc{P}_{wqo}(\cl{S}, k)$,
can now be defined as follows.
\begin{defn}\label{defn:pwqo} If $S_k$
is a w.q.o under the isomorphic embedding relation $\hookrightarrow$,
we say that $\mc{P}_{wqo}(\cl{S}, k)$ holds.
\end{defn}
Observe that $\mc{P}_{wqo}(\cl{S}, 0)$ holds iff $\cl{S}$ is a
w.q.o. under $\hookrightarrow$.  Furthermore, if $\mc{P}_{wqo}(\cl{S},
k)$ holds and $\cl{S'}$ is a subclass of $\cl{S}$, then
$\mc{P}_{wqo}(\cl{S'}, k)$ holds as well.  If $\cl{S}$ is a finite
class of structures, then $\mc{P}_{wqo}(\cl{S}, k)$ holds trivially
for each $k\in \mathbb{N}$. The next lemma provides a more interesting
example.
\begin{lemma} Let $\cl{S}$ be the class of all finite linear
orders. Then $\mc{P}_{wqo}(\cl{S}, k)$ holds for all $k \in
\mathbb{N}$.
\end{lemma}
\emph{Proof:} Fix $k$.  Let $I = (\mf{A}_i, a_i^1, \ldots, a_i^k)_{i
  \ge 1}$ be an infinite sequence of elements from $\cl{S}_k$.  Since
there are only finitely many order-types of a $k$-tuple from a linear
order, there exists an infinite subsequence $J = (\mf{B}_{j}, b_{j}^1,
\ldots, b_{j}^k)_{j \ge 1}$ of $I$ such that the order-type of $b_j^1,
\ldots, b_j^k$ in $\mf{B}_j$ is the same for all $j$.

Consider an element $(\mf{B}, b^1, \ldots, b^k)$ of $J$.  Let $b^0$
and $b^{k+1}$ be the minimum and maximum elements of $\mf{B}$.
W.l.o.g. assume $b^0 \le_{\mf{B}} b^1 \le_{\mf{B}} \ldots \le_{\mf{B}}
b^k \le_{\mf{B}} b^{k+1}$ where $\leq_{\mf{B}}$ is the linear order of
$\mf{B}$.  Then $(\mf{B}, b^1, \ldots, b^k)$ can be represented (upto
isomorphism) by a $(k+1)$-tuple $t_{\mf{B}} \in \mathbb{N}^{k{+1}}$,
where the $r^{\text{th}}$ component of $t_{\mf{B}}$ is $|\{b \mid b
\in \mathsf{U}_{\mathfrak{B}}, \, b^{r-1} \le_{\mf{B}} b \le_{\mf{B}}
b^r\}|$.  Applying Dickson's lemma to the sequence $(t_{\mf{B}_j})_{j
  \ge 1}$ there exist $p, q$ such that $p < q$ and $t_{\mf{B}_p}$ is
component-wise $\le$ $t_{\mf{B}_q}$.  Since a linear order of length
$m$ can always be embedded in a linear order of length $n$ for $n \ge
m$, it follows that $(\mf{B}_p, b_p^1, \ldots, b_p^k) \hookrightarrow
(\mf{B}_q, b_q^1, \ldots, b_q^k)$.  Hence, $\mc{P}_{wqo}(\cl{S}, k)$
holds.  \epf

As a ``mixed-example'', consider $\cl{S}$ to be the class of all
finite (undirected) paths. It is easy to see that both
$\mc{P}_{wqo}(\cl{S},0)$ and $\mc{P}_{wqo}(\cl{S},1)$ hold.  However,
$\mc{P}_{wqo}(\cl{S},k)$ fails for all $k \geq 2$. We illustrate this
failure for $k=2$. Consider the sequence $(\mf{A}_i)_{i \geq 2}$ of
structures from $\cl{S}_2$, where $\mf{A}_i=(P_i,a_i,b_i)$ and $P_i$
is a path of length $i$ with end-points $a_i$ and $b_i$.  It is easy
to check that for all $i \neq j$, $\mf{A}_i \not\hookrightarrow
\mf{A}_j$.

The following lemma provides a ``logical'' characterization of
$\mc{P}_{wqo}(\cl{S}, 0)$.  The proof can be found in Appendix 
\ref{appendix:char-of-PSC(0)}.
\begin{lemma}\label{characterization-of-PSC(0)}
Let $\cl{S}$ be a class of structures. $\mc{P}_{wqo}(\cl{S}, 0)$
holds iff ${\PS}={\PSC(0)}=\forall^*$ over $\cl{S}$.
\end{lemma}

We are now ready to state the main result of this subsection.
\begin{prop}\label{prop:P-wqo(S, k)-and-characterization-of-PSC(k)}
Let $k \in \mathbb{N}$ and $\cl{S}$ be a class of structures such 
that $\mc{P}_{wqo}(\cl{S}, k)$ holds. Then 
${\PSC}(k)=\exists^k\forall^*$ over $\cl{S}$. 
\end{prop}
\emph{Proof:} Consider $\cl{C} \in {\PSC}(k)$ over $S$.  Define
$\cl{D}$ to be the subclass of $\cl{S}_k$ consisting of all elements
$(M, a_1, \ldots, a_k)$, where $M \in \cl{C}$ and the underlying set
of $(a_1, \ldots, a_k)$ is a $k$-core of $M$ w.r.t. $\cl{C}$ over
$\cl{S}$. It follows from Definition~\ref{defn:PSC(k)} that $\cl{D}
\in {\PSC(0)}$ over $\cl{S}_k$. Since $\mc{P}_{wqo}(\cl{S}, k)$ holds,
by definition, $\mc{P}_{wqo}(\cl{S}_k, 0)$ holds.  By
Lemma~\ref{characterization-of-PSC(0)}, $\cl{D}$ is definable by a
$\forall^*$ sentence $\psi$ over $\cl{S}_k$.  We now replace each
constant in $\tau_k \setminus \tau$ that appears in $\psi$ by a fresh
variable, and existentially quantify these variables to get a
$\exists^k \forall^*$ sentence defining $\cl{C}$ over $\cl{S}$. \epf

\subsection{A logic-based combinatorial property}
For every $m \in \mathbb{N}$, $\tau$-structures $\mathfrak{A}$ and
$\mathfrak{B}$ are said to be \emph{$m$-equivalent}, denoted
$\mathfrak{A} \equiv_m \mathfrak{B}$, iff $\mathfrak{A}$ and
$\mathfrak{B}$ agree on the truth of every $FO(\tau)$ sentence of
quantifier rank $\le m$. We assume the reader is familiar with
\emph{Ehrenfeucht-Fra\"{i}ss\'{e} games} (henceforth called EF
games)~\cite{libkin,ebbinghaus-flum}.  The classical Ehrenfeuct-Fra\"{i}ss\'{e}
theorem (henceforth called EF theorem) states that $\mathfrak{A}$ and
$\mathfrak{B}$ are $m$-equivalent iff the duplicator has a winning
strategy in the $m$-round EF game between $\mathfrak{A}$ and
$\mathfrak{B}$.  

Let $k \in \mathbb{N}$ and $\cl{S}$ be a class of
structures.  Our second property, namely $\mc{P}_{logic}(\cl{S}, k)$,
can now be stated as follows.
\begin{defn}\label{defn:Plogic(S,k)}
Suppose there exists a function $f: \mathbb{N} \rightarrow \mathbb{N}$
such that for each $m \in \mathbb{N}$, for each structure $\mf{A}$ of
$\cl{S}$ and for each subset $W$ of $\mathsf{U}_{\mf{A}}$ of size at
most $k$, there exists $\mf{B} \subseteq \mf{A}$ such that (i) $\mf{B}
\in \cl{S}$, (ii) $W \subseteq \mathsf{U}_{\mf{B}}$, (iii) $|\mf{B}|
\leq f(m)$ and (iv) $\mf{B} \equiv_m \mf{A}$.  Then, we say that
$\mc{P}_{logic}(\cl{S}, k)$ holds, and call $f(m)$ a \emph{witness
  function} of $\mc{P}_{logic}(\cl{S}, k)$.
\end{defn}
Clearly, if $\mc{P}_{logic}(\cl{S}, k)$ holds and $\cl{S'}$ is a
subclass of $\cl{S}$ that is preserved under substructures over
$\cl{S}$, then $\mc{P}_{logic}(\cl{S'}, k)$ also holds.  

Revisiting
the examples of the previous subsection, we see that if $\cl{S}$ is a
finite class, then $\mc{P}_{logic}(\cl{S}, k)$ holds with $f(m)$
being the constant function that returns the size of the largest
structure in $\cl{S}$.  For linear orders, we have the following
result.
\begin{lemma} Let $\cl{S}$ be the class of all finite linear
orders. Then $\mc{P}_{logic}(\cl{S}, k)$ holds for all $k \in
\mathbb{N}$.
\end{lemma}
\emph{Proof:} Fix $m$ and $k$. Let $\mf{A} \in \cl{S}$ and let $W = \{
a_1, \ldots, a_r\}, ~r \leq k$ be a subset of $\mathsf{U}_{\mf{A}}$.
Define $f(m) = \max{\{2^m, k\}}$.  We now show that there is a linear
sub-order $\mf{B}$ of $\mf{A}$ such that $\mf{B}$ contains $W$,
$|\mf{B}| \leq f(m)$ and $\mf{B} \equiv_m \mf{A}$.

If $|\mf{A}| \leq f(m)$, choose $\mf{B} = \mf{A}$, and we are done.
Otherwise, let $\mf{B}$ be any substructure of $\mf{A}$ that contains
$W$ and is of size $f(m)$. Observe that, in this case, both
$\mf{A}$ and $\mf{B}$ are of size at least $2^m$.  It is well-known
that any two linear orders of length $\ge 2^m$ are $m$-equivalent.
Hence, $\mf{B} \equiv_m \mf{A}$, and all conditions in
Definition~\ref{defn:Plogic(S,k)} are satisfied by $\mf{B}$. \epf

Finally, if $\cl{S}$ is the class of all finite paths, it is easy to
see that $\mc{P}_{logic}(\cl{S},0)$ and $\mc{P}_{logic}(\cl{S},1)$
hold.  However, $\mc{P}_{logic}(\cl{S},k)$ fails for all $k \geq 2$.
This is because if $\mf{A}$ is a path containing two distinct
end-points, and if $W$ contains both these end-points, then $\mf{A}$
is its \emph{only} substructure \emph{in} $\cl{S}$ that contains
$W$. This precludes the existence of a uniform (i.e. independent of
$|\mf{A}|$) function $f(m)$ bounding the size of a substructure of
$\mf{A}$ containing $W$, as required in Definition~\ref{defn:Plogic(S,k)}.

The next theorem is one of the main results of this paper.  Before
stating the theorem, we note that given a class $\cl{S}$ of structures
and $n \in \mathbb{N}$, the subclass of all structures in $\cl{S}$ of
size $\le n$ is definable over $\cl{S}$ by a FO sentence in
$\exists^n\forall^*$.  We call this sentence $\xi_{\cl{S}, n}$ in the
following theorem.
\begin{theorem}\label{theorem:P-logic(S, k)-and-characterization-of-PSC(k)}
Let $\cl{S}$ be a class of structures and $k \in \mathbb{N}$ such that
$\mc{P}_{logic}(\cl{S}, k)$ holds. Then $PSC(k) = \exists^k\forall^*$
over $\cl{S}$.  More precisely, for every defining FO sentence $\phi$
in $PSC(k)$ over $\cl{S}$, there exists $p \in \mathbb{N}$ such that
$\phi$ is semantically equivalent to $\exists^k \bar{x} \forall^p
\bar{y} ~\psi|_{\bar{x}\bar{y}}$ over $\cl{S}$, where $\psi \equiv
(\xi_{\cl{S}, k+p} \rightarrow \phi)$.
\end{theorem}
\emph{Proof:} It is obvious that $\exists^k\forall^* \subseteq PSC(k)$
over $\cl{S}$.  To prove containment in the other direction, consider
$\phi$ in $PSC(k)$ over $\cl{S}$, and let $\phi$ have quantifier rank
$m$. Let $f(m)$ be the witness function of $\mc{P}_{logic}(\cl{S},
k)$.  Consider the sentence $\varphi \equiv \exists^k \bar{x}
\forall^p \bar{y} ~\psi|_{\bar{x}\bar{y}}$, where $p=f(m)$.  Since
$\phi$ is in $PSC(k)$ over $\cl{S}$, every model $\mf{A}$ of $\phi$ in
$\cl{S}$ also satisfies $\varphi$.  To see why this is so, note that
the elements of any $k$-core of $\mf{A}$ can serve as witnesses of the
existential variables in $\varphi$.  Therefore, $\phi \rightarrow
\varphi$ over $\cl{S}$.  To show $\varphi \rightarrow \phi$ over
$\cl{S}$, suppose $\mf{A}$ is a model of $\varphi$ in $\cl{S}$.  Let
$W$ be the set of witnesses in $\mf{A}$ for the $k$ existential
variables in $\varphi$.  Clearly, $|W| \le k$.  Since
$\mc{P}_{logic}(\cl{S}, k)$ holds, there exists $\mf{B} \subseteq
\mf{A}$ such that (i) $\mf{B} \in \cl{S}$, (ii) $W \subseteq
\mathsf{U}_{\mf{B}}$, (iii) $|\mf{B}| \leq f(m)=p$, and (iii) $\mf{B}
\equiv_m \mf{A}$.  Since $\mf{A} \models \varphi$, by instantiating
the universal variables in $\varphi$ with the elements of
$\mathsf{U}_{\mf{B}}$, we have $\mf{B} \models \phi$.  Since the
quantifier rank of $\phi$ is $m$ and $\mf{B} \equiv_m \mf{A}$, it
follows that $\mf{A} \models \phi$. Therefore, $\phi$ is semantically
equivalent to $\varphi$ over \cl{S}.  This proves the theorem.  \epf

{\bfseries Remark:} Suppose $\mf{A} \in \cl{S}$ is a model of
$\phi$, and $\bar{a}=(a_1,a_2,\ldots,a_k) \in \mathsf{U}_{\mf{A}}^k$.
Define $\phi'(\bar{x}) \equiv \forall^p
\bar{y}~\psi|_{\bar{x}\bar{y}}$.  It is easy to see that $\mf{A}
\models \phi'(\bar{a})$ iff the underlying-set
$\{a_1,a_2,\ldots,a_k\}$ is a $k$-core of $\mf{A}$ w.r.t. $\phi$.
Thus, the $k$-cores of $\phi$ are defined by the FO formula
$\phi'(\bar{x})$ if $\mc{P}_{logic}(\cl{S}, k)$ holds.
\vspace{2pt}

The witness function $f(m)$ in the definition of
$\mc{P}_{logic}(\cl{S}, k)$ may not be computable, in general.  By
requiring $f$ to be \emph{computable} in Definition~\ref{defn:Plogic(S,k)},
we obtain an effective version of $\mc{P}_{logic}(\cl{S}, k)$, which
we call $\mc{P}_{logic}^{comp}(\cl{S}, k)$.  Note that for all
examples considered so far where $\mc{P}_{logic}(\cl{S}, k)$ holds, we
actually gave closed form expressions for $f(m)$.  Hence
$\mc{P}_{logic}^{comp}(\cl{S}, k)$ also holds for these classes.  We
will soon see a class $\cl{S}$ that is in $\mc{P}_{logic}(\cl{S}, k)$,
but not in $\mc{P}^{comp}_{logic}(\cl{S}, k)$.
\vspace{2pt}

The following is an important corollary of
Theorem~\ref{theorem:P-logic(S, k)-and-characterization-of-PSC(k)}.
\begin{corollary}\label{theorem:P-logic(S, k)-and-characterization-of-PSC(k):effective-version}
Let $\cl{S}$ be a class of structures and $k \in \mathbb{N}$ such that
$\mc{P}^{comp}_{logic}(\cl{S}, k)$ holds. For every $\phi$ in $PSC(k)$
over $\cl{S}$, the translation to a semantically equivalent (over
$\cl{S}$) $\exists^k\forall^*$ sentence $\varphi$ is effective.
\end{corollary}

\section{Relations between properties}\label{section:relations}
We begin by comparing the classes for which $\mc{P}_{wqo}(\cl{S}, k)$
hold with those for which $\mc{P}_{logic}(\cl{S}, k)$ hold.
Surprisingly, it turns out that $\mc{P}_{wqo}(\cl{S}, k)$ implies
$\mc{P}_{logic}(\cl{S}, k)$!
\vspace{2pt}
\begin{prop}\label{prop:pwqo-implies-plogic}
For each class $\cl{S}$ of structures and each $k \in \mathbb{N}$, if
$\mc{P}_{wqo}(\cl{S}, k)$ holds, then $\mc{P}_{logic}(\cl{S}, k)$ holds
as well.
\end{prop}
\emph{Proof}: We give a proof by contradiction.  Suppose, if possible,
$\mc{P}_{wqo}(\cl{S}, k)$ holds but $P_{logic}(\cl{S}, k)$ fails.  By
Definition~\ref{defn:Plogic(S,k)}, there exists $m \in \mathbb{N}$
such that for all $p \in \mathbb{N}$, there exists a structure
$\mf{A}_p$ in $\cl{S}$ and a set $W_p$ of at most $k$ elements of
$\mf{A}_p$ such that for any substructure $\mf{B}$ of $\mf{A}_p$, we
have $\big((\mf{B} \in \cl{S}) \wedge (W_p \subseteq
\mathsf{U}_{\mf{B}}) \wedge (|\mf{B}| \leq p)\big) \rightarrow (\mf{B}
\not\equiv_m \mf{A}_p)$.

For each $p \geq 1$, fix the structure $\mf{A}_p$ and the set $W_p$
that satisfy these properties.  Now, let $\bar{a}_p$ be any $k$-tuple
such that the components of $\bar{a}_p$ are exactly the elements of
$W_p$. Let $\mf{A}'_p$ be the $\tau_k$-structure $(\mf{A}_p,
\bar{a}_p)$ in $\cl{S}_k$.  Consider the sequence $(\mf{A}'_i)_{i \geq
  1}$.  Since $\mc{P}_{wqo}(\cl{S}, k)$ holds, $\cl{S}_k$ is a
w.q.o. under $\hookrightarrow$.  Therefore, there exists an infinite
sequence of indices $I = (i_1, i_2, \ldots)$ such that $i_1 < i_2 <
\ldots$ and $\mf{A}'_{i_1} \hookrightarrow \mf{A}'_{i_2}
\hookrightarrow \ldots$.  Let $\Delta$ be the set of all equivalence
classes of the $\equiv_m$ relation over the class of all
$\tau$-structures.  Given $m$ and $\tau$, $\Delta$ is clearly a finite
set.  Therefore, there exists an infinite subsequence $J$ of $I$ with
indices $j_1 < j_2 < \ldots$ such that (i) $\mf{A}'_{j_1}
\hookrightarrow \mf{A}'_{j_2} \hookrightarrow \ldots$, and (ii) the
corresponding $\tau$-reducts $\mf{A}_{j_1}, \mf{A}_{j_2}, \ldots$ are
$m$-equivalent.  Let $r = |\mf{A}'_{j_1}|$ (which is the same as
$|\mf{A}_{j_1}|$) and let $n > 1$ be an index such that $j_n \geq r$.
Then $\mf{A}'_{j_1} \hookrightarrow \mf{A}'_{j_n}$ and $\mf{A}_{j_1}
\equiv_m \mf{A}_{j_n}$.  Fix an embedding $\imath: \mf{A}'_{j_1}
\hookrightarrow \mf{A}'_{j_n}$ in $\cl{S}_k$.  We abuse notation and
denote the induced embedding of $\imath$ on the $\tau$-reducts also by
$\imath:\mf{A}_{j_1} \hookrightarrow \mf{A}_{j_n}$ in $\cl{S}$.

Let $\mf{B}$ be the image of $\mf{A}_{j_1}$ under $\imath$.  Then
$\mf{B}$ has the following properties: (i) $\mf{B} \in \cl{S}$, since
$\mf{A}_{j_1} \in \cl{S}$ and $\cl{S}$ is closed under isomorphisms,
(ii) $W_{j_n} \subseteq \mathsf{U}_\mf{B}$, since $\imath:
(\mf{A}_{j_1}, \bar{a}_{j_1}) \hookrightarrow (\mf{A}_{j_n},
\bar{a}_{j_n})$ and the components of $\bar{a}_{j_n}$ are exactly the
elements of $W_{j_n}$, (iii) $|\mf{B}|= r \leq j_n$, and (iv) $\mf{B}
\equiv_m \mf{A}_{j_n}$.  This contradicts the property of
$\mf{A}_{j_n}$ stated at the outset, completing the proof.\epf
\vspace{2pt}

Given the above result, it is natural to ask whether $\mc{P}_{wqo}(\cl{S}, k)$
implies $\mc{P}_{logic}^{comp}(\cl{S}, k)$ as
well. Proposition~\ref{prop:pwqo-does-not-imply-p-comp-logic}
provides a strong negative answer to this question.
\begin{prop}\label{prop:pwqo-does-not-imply-p-comp-logic}
There exists a class $\cl{S}$ of structures for which
$\mc{P}_{wqo}(\cl{S}, k)$, and hence $\mc{P}_{logic}(\cl{S}, k)$,
holds for every $k \in \mathbb{N}$, but $\mc{P}^{comp}_{logic}(\cl{S},
0)$ fails, and hence $\mc{P}^{comp}_{logic}(\cl{S}, k)$ fails for
every $k \in \mathbb{N}$.
\end{prop}
\emph{Proof:} Recall that the set of all computable functions from
$\mathbb{N}$ to $\mathbb{N}$ is countable.  Fix an enumeration $f_0,
f_1, f_2, f_3, \cdots$ of the computable functions.  Now define a
class $\cl{S}$ of words over the alphabet $\Sigma=\{a,b,c\}$ as
follows: $\cl{S}=\{w_i=a b^i a c^{f_i(i+3)} \mid i \in \mathbb{N}\}$.
We show in Section~\ref{section:words-and-trees} that the class
$\Sigma^*$ of all finite words satisfies $\mc{P}_{wqo}(\Sigma^*, k)$
for every $k \in \mathbb{N}$.  It follows that $\mc{P}_{wqo}(\cl{S},
k)$ also holds for every $k \in \mathbb{N}$.  Now we prove by
contradiction that $\mc{P}^{comp}_{logic}(\cl{S}, 0)$ fails.  If
possible, suppose $\mc{P}^{comp}_{logic}(\cl{S}, 0)$ holds, and let
$f_r(m)$ be a computable witness function for
$\mc{P}^{comp}_{logic}(\cl{S}, 0)$.  Consider the word $w_r = a b^r a
c^{f_r(r+3)}$. By definition of $\mc{P}^{comp}_{logic}(\cl{S}, 0)$,
there exists a subword $w' \in \cl{S}$ such that $|w'| \leq f_r(r+3)$
and $w' \equiv_{r+3} w_r$.  Since $w' \equiv_{r+3} w_r$, it is easy to
see that $w'$ and $w_r$ must agree on the first $r+2$ letters.
However, since $w'$ must also be in $\cl{S}$, by the definition of
$\cl{S}$, we must choose $w' = w_r$. This contradicts $|w'| \leq
f_r(r+3)$.  \epf

{\bfseries Remark:} If $\cl{S}$ is recursively enumerable, the witness
function in Definition~\ref{defn:Plogic(S,k)} lies in the second level
of the Turing hierarchy.  See Appendix \ref{appendix:turing-hierarchy} for a
proof.

The following proposition, along with
Proposition~\ref{prop:pwqo-does-not-imply-p-comp-logic} shows that
$\mc{P}_{logic}^{comp}(\cl{S}, k)$ and $\mc{P}_{wqo}(\cl{S}, k)$ are
incomparable.
\begin{prop}\label{prop:p-comp-logic-does-not-imply-pwqo}
There exists a class $\cl{S}$ of structures for which
$\mc{P}_{logic}^{comp}(\cl{S},k)$ holds for all $k \in \mathbb{N}$, but 
$\mc{P}_{wqo}(\cl{S}, 0)$ fails, and hence $\mc{P}_{wqo}(\cl{S}, k)$
fails for all $k \in \mathbb{N}$.
\end{prop}
\ti{Proof sketch:} Let $C_n$ (respectively, $P_n$) denote an
undirected cycle (respectively, path) of length $n$. Let $mP_n$ denote
the disjoint union of $m$ copies of $P_n$. Let $H_n = \bigsqcup_{i =
  0}^{i = 3^n} nP_i$ and $G_n = C_{3^n} \sqcup H_n$, where $\sqcup$
denotes disjoint union.  Now consider the class $\cl{S}$ of undirected
graphs given by $\cl{S} = \cl{S}_1 \cup \cl{S}_2$, where $\cl{S}_1 =
\{H_n \mid n \ge 1\}$ and $\cl{S}_2 = \{ G_n \mid n \ge 1\}$.  That
$\mc{P}_{wqo}(\cl{S}, 0)$ fails is easily seen by considering the
sequence $(G_n)_{n \ge 1}$, and noting that $C_{3^n}$ cannot embed in
$C_{3^m}$ unless $m = n$.  The proof that
$\mc{P}_{logic}^{comp}(\cl{S}, k)$ holds for all $k$ is deferred to
Appendix \ref{appendix:cycles+paths}.\epf

\vspace{2pt}
Towards a comparison of classes satisfying the various properties
defined above, we consider the following eight natural collections of
classes and study the relations between them.

\begin{tabular}{lll}
1) & $\Gamma_{wqo}^0$ & $=\, \{ \cl{S} \mid \mc{P}_{wqo}(\cl{S}, 0)
\mbox{~holds~}\}$\\
2) & $\Gamma_{wqo}^*$ & $=\, \{ \cl{S} \mid \forall k\;
\mc{P}_{wqo}(\cl{S}, k) \mbox{~holds~}\}$\\
3) & $\Gamma_{logic}^0$ & $=\, \{ \cl{S} \mid \mc{P}_{logic}(\cl{S},
0) \mbox{~holds~}\}$\\
4) & $\Gamma_{logic}^*$ & $=\, \{ \cl{S} \mid \forall k\;
\mc{P}_{logic}(\cl{S}, k) \mbox{~holds~}\}$\\
5) & $\Gamma_{comp}^0$ & $=\, \{ \cl{S} \mid
\mc{P}_{logic}^{comp}(\cl{S}, 0) \mbox{~holds~}\}$\\
6) & $\Gamma_{comp}^*$ & $=\, \{ \cl{S} \mid \forall k\;
\mc{P}_{logic}^{comp}(\cl{S}, k) \mbox{~holds~}\}$\\
7) & $\Gamma^0$ & $=\, \{ \cl{S} \mid PS=PSC(0)=\forall^*
\mbox{~over~} \cl{S} \}$\\
8) & $\Gamma^*$ & $=\, \{ \cl{S} \mid \forall k\;
PSC(k)=\exists^k\forall^* \mbox{~over~} \cl{S} \}$\\
\end{tabular}

Note that $\Gamma^0$ (repectively, $\Gamma^*$) is the collection of
all classes of finite structures over which the {\lt} theorem
(respectively, $PSC(k) = \exists^k\forall^*$) holds.  Using
$\subseteq$ to denote containment for collections of classes, it is
trivial to see that $\Gamma^* \subseteq \Gamma^0$, $\Gamma_{wqo}^*
\subseteq \Gamma_{wqo}^0$, $\Gamma_{logic}^* \subseteq
\Gamma_{logic}^0$, $\Gamma_{comp}^* \subseteq \Gamma_{comp}^0$,
$\Gamma_{comp}^* \subseteq \Gamma_{logic}^*$ and 
$\Gamma_{comp}^0 \subseteq \Gamma_{logic}^0$.
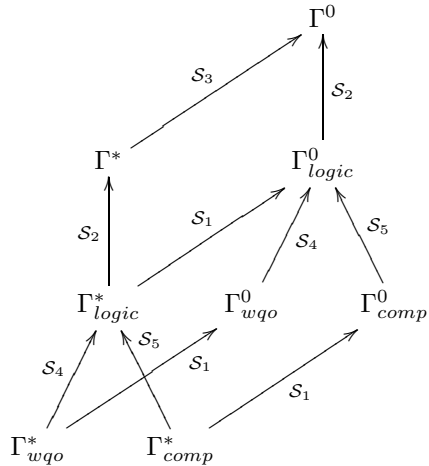
\[
      \xymatrix@!=0.25pc{
      & & & & \Gamma^0 & \\
      & & & & & \\
      & \Gamma^*\ar[uurrr]^(0.5){\cl{S}_3} & & & \Gamma_{logic}^0\ar[uu]_(0.5){\cl{S}_2} & \\
      & & & & & \\
      & \Gamma_{logic}^* \ar[uurrr]^(0.5){\cl{S}_1}\ar[uu]^(0.5){\cl{S}_2} & & \Gamma_{wqo}^0\ar[uur]_(0.5){\cl{S}_4} & & \Gamma_{comp}^0 \ar[uul]_(0.5){\cl{S}_5} \\
      & & & & & & \\
      \Gamma_{wqo}^* \ar[uurrr]_(0.70){\cl{S}_1} \ar[uur]^(0.5){\cl{S}_4} & & \Gamma_{comp}^* \ar[uurrr]_(0.5){\cl{S}_1} \ar[uul]_(0.7){\cl{S}_5} & & &}
\]

The ``Hasse'' diagram $\mc{D}$ above depicts all containment relations
between the above eight collections.  Every directed arrow represents
a ``primary'' containment.  Some of these containments have already
been discussed above.  The remaining follow from Theorem
\ref{theorem:P-logic(S, k)-and-characterization-of-PSC(k)} and from
Proposition \ref{prop:pwqo-implies-plogic}.  Every pair of unordered
collections in the diagram represents incomparable collections.  We
also annotate every directed arrow with a label that refers to an
example demonstrating the \emph{strictness} of the containment.  The
list of examples used to show strict containments is as follows.
\begin{itemize}
\item $\cl{S}_1$ is the class of all undirected paths. 
\item $\cl{S}_2$ is the class of all undirected cycles.
\item $\cl{S}_3$ is the class of all undirected graphs that are disjoint unions of paths.
\item $\cl{S}_4$ is the class of structures constructed in the proof of Proposition~\ref{prop:p-comp-logic-does-not-imply-pwqo}.
\item $\cl{S}_5$ is the class of words constructed in the proof of
  Proposition~\ref{prop:pwqo-does-not-imply-p-comp-logic}.
\end{itemize}

For each example in the above list, we indicate below all the
lowest/minimal collections in $\mc{D}$ that contains it, and all the
highest/maximal collections in $\mc{D}$ that does not contain it. Due
to lack of space, proofs of these inclusions/non-inclusions are
deferred to Appendix \ref{appendix:relations}.  We discuss only the
inclusions/non-inclusions of $\cl{S}_2$ in detail below.

\begin{itemize}
\item $\cl{S}_1 \in \Gamma^0_{wqo},\, \cl{S}_1 \in \Gamma^0_{comp},\, \cl{S}_1 \in \Gamma^*,\,
  \cl{S}_1 \notin \Gamma^*_{logic}$.
\item $\cl{S}_2 \in \Gamma^*,\, \cl{S}_2 \notin \Gamma^0_{logic}$.
\item $\cl{S}_3 \in \Gamma^0_{wqo},\, \cl{S}_3 \in \Gamma^0_{comp},\,
  \cl{S}_3 \notin \Gamma^*$.
\item $\cl{S}_4 \in \Gamma^*_{comp},\, \cl{S}_4 \notin \Gamma^0_{wqo}$.
\item $\cl{S}_5 \in \Gamma^*_{wqo},\, \cl{S}_5 \notin \Gamma^0_{comp}$.
\end{itemize}

\begin{lemma}\label{lemma:cycles} 
The class $\cl{S}_2$ belongs to $\Gamma^* \setminus \Gamma_{logic}^0$.
\end{lemma}
\emph{Proof:} For any cycle $\mf{A} \in \cl{S}_2$, the only
substructure of $\mf{A}$ in $\cl{S}_2$ is $\mf{A}$ itself. It follows
that $\cl{S}_2 \notin \Gamma_{logic}^0$.  We now show that $\cl{S}_2
\in \Gamma^*$. Firstly, observe that \emph{any} subclass of $\cl{S}_2$
is in ${\PS}$ over $\cl{S}_2$. So, it suffices to show that $\cl{S}_2$
belongs to $\Gamma^0$.  Towards this, let $\phi$ be in $PS$ over
$\cl{S}_2$, and suppose the quantifier rank of $\phi$ is $m$.  It is
well-known that any two cycles of sizes $\ge p = 2^m$ are
$m$-equivalent. Then either (i) all cycles of size $\ge p$ are models
of $\phi$ or (ii) all models of $\phi$ in $\cl{S}_2$ have sizes $< p$.

In case (i), we define $\phi'=\forall^p \bar{y} ~\psi|_{\bar{y}}$
where $\psi \equiv \xi_{\cl{S}_2, p} \rightarrow \phi$ and
$\xi_{\cl{S}_2, p}$ is as in Theorem~\ref{theorem:P-logic(S,
  k)-and-characterization-of-PSC(k)}.  It is easy to verify that (a)
all cycles of size $> p$ are models of $\phi'$, and (b) a cycle of
length $\le p$ is a model of $\phi$ iff it is a model of
$\phi'$. Therefore, $\phi$ is equivalent to $\phi'$ over $\cl{S}_2$.

In case (ii), let $X$ be the finite set of the sizes of all models of
$\phi$ in $\cl{S}_2$.  It is trivial to see that there exists a
quantifier-free formula $\xi'_X(y_1,y_2,\ldots,y_p)$ which asserts the
following: if the substructure induced by $y_1, y_2, \ldots, y_p$ is a
cycle, then its size belongs to $X$.  From here, it is easy to check
that $\phi$ is equivalent to $\forall^p \bar{y}
\xi'_X(y_1,y_2,\ldots,y_p)$ over $\cl{S}_2$.

In either case, we have shown that $\phi$ is equivalent to a
$\forall^*$-sentence over $\cl{S}_2$. This proves that $\cl{S}_2$
belongs to $\Gamma^*$.  \epf

\section{Words and trees over a finite alphabet}\label{section:words-and-trees}

Given a finite alphabet $\Sigma$, let ${\words}(\Sigma)$ and
${\trees}(\Sigma)$ denote the set of all finite words and finite
trees, respectively, over $\Sigma$.  In this section, we show that
$\mc{P}_{wqo}(\cl{S}, k)$ and $\mc{P}^{comp}_{logic}(\cl{S}, k)$ hold
for every $k \in \mathbb{N}$, in each of the two cases, $\cl{S} =
{\words}(\Sigma)$ and $\cl{S} = {\trees}(\Sigma)$.

Various representations of trees have been used in the literature.  In
this paper, we use the following poset-theoretic representation.  A
tree is a finite poset $P = (A, \leq)$ with a unique minimal element
(called root), and for every $a, b, c \in A$, $\big((a \leq c)
~\wedge~ (b \leq c)\big) \rightarrow \big(a \leq b ~\vee~ b \leq
a\big)$.  Informally, the Hasse diagram of $P$ is an (inverted) tree
with $p$ connected to $c$ for every parent $p$ and its child $c$.  A
tree over $\Sigma$ (henceforth called a \emph{$\Sigma$-tree}) is a
pair $(P, f)$ where $P = (A, \leq)$ is a tree and $f: A \rightarrow
\Sigma$ is a labeling function.  The elements of $A$ are also called
nodes (or elements) of the $\Sigma$-tree $((A, \leq), f)$.  We use
$root_s$ to denote the root of a $\Sigma$-tree $s$.  In the special
case where the underlying poset is a linear order, a $\Sigma$-tree is
called a $\Sigma$-word.

Let $\tau$ be the vocabulary $\{\leq\} \cup \{Q_a \mid a \in
\Sigma\}$, where $\leq$ is a binary predicate and each $Q_a$ is a
unary predicate.  A $\Sigma$-tree $t = ((A, \leq_t), f)$ can be
naturally represented as a structure $\mf{A}_t$ over $\tau$.
Specifically, $\mathsf{U}_{\mf{A}_t} = A$, the interpretation
$\leq^{\mf{A}_t}$ is the same as $\leq_t$, and for every $a \in
\Sigma$, $Q_a^{\mf{A}_t}$ represents the set of all elements of $A$
labeled $a$ by $f$.  For the special case of a $\Sigma$-word $w$, we
let $A$ be $\{1,\ldots |w|\}$, and use $w[j]$ to denote the letter at
the $j^{th}$ position of $w$.  A $\Sigma$-forest ${\f}$ is a (finite)
disjoint union of $\Sigma$-trees.  We use the disjoint union of the
$\tau$-structures representing the $\Sigma$-trees in ${\f}$ to represent
the $\Sigma$-forest ${\f}$.  For clarity of exposition, we abuse
notation and use $t$ to denote both a $\Sigma$-tree and the
corresponding $\tau$-structure $\mf{A}_t$, when it is clear from
the context which of these we are referring to. 

Given two $\Sigma$-trees $\s = ((A_{\s}, \leq_{\s}), f_{\s})$ and
$\myt = ((A_t, \leq_t), f_t)$, and an element $e$ of $\s$, the
\emph{join of $t$ to $\s$ at $e$}, denoted $\s \cdot_e t$, is the
$\Sigma$-tree defined upto isomorphism as follows. Let $t' = ((A_t',
\leq_t'), f_t')$ be an isomorphic copy of $t$ such that $A_{\s} \cap
A_t' = \emptyset$.  Then $\s \cdot_e t$ is the $\Sigma$-tree $((A,
\leq), f)$, where (i) $A = A_{\s} \cup A_t'$ (ii) $f$ is: $f_{\s}$ on
$A_s$ and $f_t'$ on $A_t'$, and (iii) $\leq \, = \, \leq_{\s} \cup
\leq_t' \cup ~\{ (c, d) \mid c \leq_{\s} e,~ c \in A_{\s}, ~d \in
A_t'\}$.  Given $\Sigma$-trees $\s$ and $t$ as above, we say $t$ is a
\emph{subtree} of $\s$ if (i) $A_t \subseteq A_{\s}$, (ii) $\leq_t$ is
the restriction of $\leq_{\s}$ to $A_t$, and (iii) $f_t$ is the
restriction of $f_{\s}$ to $A_t$.  If $\s$ is a $\Sigma$-word, a
subtree of $\s$ is also called a \emph{subword}.  Given a
$\Sigma$-tree $\s$, a $\Sigma$-forest $\f = \sqcup_{i \in \{1, \ldots,
  n\}} t_i$ and an element $e$ of $\s$, the \emph{the join of $\f$ to
  $\s$ at $e$}, denoted $\s \cdot_e \f$, is the $\Sigma$-tree defined
upto isomorphism as $((((\s \cdot_e t_1) \cdot_e t_2) \cdots) \cdot_e
t_n)$.

\subsection{$\mc{P}_{wqo}(\cl{S}, k)$ for words and trees}\label{subsection:pwqo-over-words-and-trees}

The primary result of this subsection is the following.
\vspace{2pt}
\begin{prop}\label{prop:words-and-trees-satisfy-pwqo(S, k)}
Given a finite alphabet $\Sigma$, for every $k \in \mathbb{N}$,
$\mc{P}_{wqo}({\words}(\Sigma), k)$ and
$\mc{P}_{wqo}({\trees}(\Sigma), k)$ hold.
\end{prop}
\vspace{2pt} Our proof goes via an alternative definition of
$\mc{P}_{wqo}(\cl{S}, k)$.  Given a vocabulary $\tau$ and a unary
predicate $R \not\in \tau$, let $\tau_R$ denote the vocabulary $\tau
\cup \{R\}$.  For a class $\cl{S}$ of $\tau$-structures, we denote by
$\cl{S}^k$ the class of all $\tau_R$-structures $(\mf{A}, R^{\mf{A}})$
s.t. $\mf{A} \in \cl{S}$ and $|R^{\mf{A}}| \leq k$.  The following
lemma shows that $\mc{P}_{wqo}(\cl{S}^k, 0)$ serves as an alternative
definition of $\mc{P}_{wqo}(\cl{S},k)$.
\vspace{2pt}
\begin{lemma}\label{lemma:towards-two-equivalent-defns-of-P-wqo}
For every class $\cl{S}$ of structures and $k \in \mathbb{N}$,
$\mc{P}_{wqo}(\cl{S}, k)$ holds iff $\mc{P}_{wqo}(\cl{S}^k, 0)$ holds.
\end{lemma}
\vspace{2pt} It is trivial to see that $\mc{P}_{wqo}(\cl{S},k)$
implies $\mc{P}_{wqo}(\cl{S}^k, 0)$. See Appendix 
\ref{appendix:alt-defn-of-Pwqo} for a proof of converse.

Proposition~\ref{prop:words-and-trees-satisfy-pwqo(S, k)} can now be
proved using the above definition of $\mc{P}_{wqo}(\cl{S}, k)$.  We
first consider the class ${\words}(\Sigma)$. Let $(w_i, R_i)_{i \ge
  1}$ be an infinite sequence of structures from ${\words}(\Sigma)^k$.
Define $\widetilde{\Sigma} = \{\tilde{a} \mid a \in \Sigma\}$, where
$\tilde{a} \not\in \Sigma$ for each $a \in \Sigma$.  Let $v_i$ be a
word over $\Sigma \cup \tilde{\Sigma}$ s.t. (i) $|v_i| = |w_i|$, and
(ii) for $j$ ranging over the positions of $w_i$, $v_i[j] = w_i[j]$ if
$j \notin R_i$, and $v_i[j] = \widetilde{w_i[j]}$ otherwise.  Now consider
the sequence of words $v_1, v_2, \ldots$ over $\Sigma \cup
\widetilde{\Sigma}$.  By Higman's Lemma, there exist $i, j$ s.t.  $i <
j$ and $v_i$ is a subword of $v_j$.  It follows that $(w_i, R_i)
\hookrightarrow (w_j, R_j)$, when viewed as $\tau_R$-structures.  The
proof for the class ${\trees}(\Sigma)$ is similar, and uses Kruskal's
tree theorem instead of Higman's lemma.\epf

\subsection{$\mc{P}_{logic}^{comp}(\cl{S}, k)$ for words and trees}\label{subsection:words-and-trees-satisfy-plogic(S, k)}

We now prove the counterpart of
Proposition~\ref{prop:words-and-trees-satisfy-pwqo(S, k)} for
$\mc{P}_{logic}^{comp}(\cl{S}, k)$.
\begin{prop}\label{prop:words-and-trees-satisfy-plogic(S, k)}
Given a finite alphabet $\Sigma$, for every $k \in \mathbb{N}$,
$\mc{P}_{logic}^{comp}({\words}(\Sigma), k)$  and
$\mc{P}_{logic}^{comp}({\trees}(\Sigma), k)$ hold.
\end{prop}
Since the result for ${\trees}(\Sigma)$ subsumes that for
${\words}(\Sigma)$, we discuss the proof only for ${\trees}(\Sigma)$.

The proof of Proposition~\ref{prop:words-and-trees-satisfy-plogic(S,
k)} makes crucial use of a helper lemma.  To help explain the lemma,
we introduce some notation.  Given an alphabet $\Sigma$ and
$m \in \mathbb{N}$, let $\Delta(m, \Sigma)$ denote the set of all
equivalence classes of the $\equiv_m$ relation over
${\trees}(\Sigma)$.  Let
$f: \mathbb{N} \times \mathbb{N} \rightarrow \mathbb{N}$ be a function
such that $f(m, |\Sigma|)$ gives the number of equivalence classes of
$\equiv_m$ over ${\trees}(\Sigma)$.  Given a $\Sigma$-tree $s$ and an
element $a$ in $s$, we use ${\s}_{\ge a}$ to denote the subtree of
$\s$ induced by elements $b$ that satisfy $\s \models (a \leq b)$.  We
define ${\s}_{\not\ge a}$ and ${\s}_{\not> a}$ in an analogous
manner. Intuitively, ${\s}_{\ge a}$ is the subtree of $\s$ rooted at
$a$, ${\s}_{\not\ge a}$ is the subtree obtained by removing the
subtree rooted at $a$, and ${\s}_{\not> a}$ is the subtree of $\s$
obtained by removing all descendants of $a$.  We also define
the \emph{height} of a $\Sigma$-tree to be the length of the longest
chain in the underlying poset.  Furthermore, we say that a
$\Sigma$-tree has degree $\le d$ if every node (element) in the tree
has $\le d$ children.  The helper lemma can now be stated as follows.
\begin{lemma}\label{lemma:bounding-the-degree-and-height-given-W}
Let $s$ be a $\Sigma$-tree and $W$ be a subset of elements of $s$
s.t. $|W| \le k$.  For every $m \in \mathbb{N}$, the following
hold.
\begin{enumerate}
\item[(a)] There exists a subtree $t_1$ of $s$ s.t. (i)
$t_1$ contains all elements of $W$, (ii) $t_1$ has degree at most
$d(m, |\Sigma|)$, and (iii) $t_1 \equiv_m s$, where
$d(m, n) = (m+k)\cdot f(m, n)$ for every $m, n \in \mathbb{N}$.
\item[(b)] There exists a subtree $t_2$ of $s$ s.t. 
(i) $t_2$ contains all elements of $W$, (ii) $t_2$ has height at most
$h(m, |\Sigma|)$, and (iii) $t_2 \equiv_m s$, where $h(m, n) =
k^2 \cdot (f(m, 3f(m, n)) + 2) + f(m, n) + 1$ for every $m,
n \in \mathbb{N}$.
\end{enumerate}
\end{lemma}
Given Lemma~\ref{lemma:bounding-the-degree-and-height-given-W}, it is
easy to see that for every $m \in \mathbb{N}$, there exists a a
subtree $t$ of $s$ s.t. (i) $t$ contains all elements of $W$, (ii) the
size of $t$ is bounded by a computable function on $m$ and $|\Sigma|$
that uses $f(\cdot,\cdot)$ as an oracle, and (iii) $t \equiv_m s$.
Since $f(\cdot, \cdot)$ is indeed a computable function, and 
since every subtree of $s$ is also a $\Sigma$-tree, this proves
Proposition~\ref{prop:words-and-trees-satisfy-plogic(S, k)}.

To prove Lemma~\ref{lemma:bounding-the-degree-and-height-given-W}, we
need a few additional auxiliary lemmas.  An easy but important one
among them is the following composition lemma.
\begin{lemma}\label{lemma:composition-lemma-for-trees}
Let ${\s}_i$ be a non-empty $\Sigma$-tree containing element $a_i$,
and ${\f}_i$ be a non-empty $\Sigma$-forest containing element $b_i$,
for $i \in \{1, 2\}$.  Let ${\myr}_i = {\s}_i \cdot_{a_i} {\f}_i$ for
$i \in \{1, 2\}$.  Suppose $({\s}_1, a_1) \equiv_m ({\s}_2, a_2)$.
Then the following hold.
\begin{enumerate}
\item If $({\f}_1, b_1) \equiv_m ({\f}_2, b_2)$, then $({\myr}_1, a_1, b_1)  \equiv_m ({\myr}_2, a_2, b_2)$.  
It follows that (i) $({\myr}_1,  a_1) \equiv_m ({\myr}_2, a_2)$ (ii) $({\myr}_1, b_1) \equiv_m
  ({\myr}_2, b_2)$ and (iii) ${\myr}_1 \equiv_m {\myr}_2$.
\item If ${\f}_1 \equiv_m {\f}_2$, then $({\myr}_1, a_1) \equiv_m ({\myr}_2,  a_2)$. It follows that ${\myr}_1 \equiv_m {\myr}_2$.
\end{enumerate}
\end{lemma}
\ti{Proof Sketch}: 
For part (1), the winning strategy for the duplicator in the $m$-round
EF game between $({\myr}_1, a_1, b_1)$ and $({\myr}_2, a_2, b_2)$ is
the composition of the winning strategies for the duplicator in the
$m$-round games between $({\s}_1, a_1)$, $({\s}_2, a_2)$ and $({\f}_1,
b_1)$, $({\f}_2, b_2)$.  A similar argument works for part (2) as
well. See Appendix \ref{appendix:tree-composition-lemma}.\epf\\ Note that
composition results of this kind were first studied by Feferman and
Vaught, and subsequently by others (see~\cite{makowsky} for a survey).

Lemma~\ref{lemma:bounding-the-degree-and-height-given-W}(a) can
now be proved as follows.
\ti{Proof}: 
Let $d = (m + k) \cdot f(m, |\Sigma|)$, where $m$ and $k$ are as in
the statement of
Lemma~\ref{lemma:bounding-the-degree-and-height-given-W}.  For an
element $a$ in $\s$, let ${\Children}(a)$ denote the set of all its
children in $\s$.  If $|{\Children}(a)| \leq d$ for every $a$ in $\s$,
we choose $t_1$ to be the same as $s$, and we are done.  Otherwise,
suppose $|{\Children}(a)| > d$.  Let $W_a = \{b \in
{\Children}(a) \mid s_{\ge b}~\text{contains an element
of}~W\}$. Observe that $|W_a| \leq k$. For every
$\delta \in \Delta(m, \Sigma)$, we define the following subsets of
${\Children}(a)$:
\begin{itemize}
\item $C(a, \delta)$ $=$ $\{b \in {\Children}(a) \mid~
\text{the}~\equiv_m~\text{class of}~{\s}_{\ge b}~\text{is}~\delta\}$. 
\item $D(a, \delta) = C(a, \delta)$, if $|C(a, \delta)| < m + k$;
otherwise, $D(a, \delta)$ is any subset of $C(a, \delta)$ s.t.
$(W_a \cap C(a, \delta)) \subseteq D(a, \delta)$ and $|D(a, \delta)| =
m + k$.
\end{itemize}
It is easy to check that the forest defined by the subtrees of $\s$
rooted at $C(a, \delta)$ is $m$-equivalent to the forest defined by
the subtrees rooted at $D(a, \delta)$.  Formally, if ${\f}_\delta
= \bigsqcup_{b \in C(a, \delta)}~{\s}_{\ge b}$ and ${\g}_\delta
= \bigsqcup_{b \in D(a, \delta)}~{\s}_{\ge b}$, then
${\f}_\delta \equiv_m {\g}_\delta$.  Therefore, if ${\f}
= \bigsqcup_{\delta \in \Delta(m, \Sigma)}~ {\f}_\delta$ and ${\g}
= \bigsqcup_{\delta \in \Delta(m, \Sigma)}~ {\g}_\delta$, we have
${\f} \equiv_m {\g}$.  By
Lemma \ref{lemma:composition-lemma-for-trees}, ${\s} =$
$\big({\s}_{\not> a} \cdot_a {\f}\big) \equiv_m \big({\s}_{\not>
a} \cdot_a {\g}\big)$ $=$ $s_1$, say.  Thus ${\s}_1$ contains $W$ and
the element $a$ has at most $d$ children in ${\s}_1$.  The count of
elements in ${\s}_1$ having $> d$ children is therefore strictly less
than the corresponding number in ${\s}$.  Since $\s$ has finitely many
elements, by repeating the above argument, we obtain a subtree $t_1$
of $\s$, as required in
Lemma~\ref{lemma:bounding-the-degree-and-height-given-W}(a).\epf
\vspace{2pt}

To prove Lemma~\ref{lemma:bounding-the-degree-and-height-given-W}(b),
we need a few additional auxiliary lemmas.  As before, let $\s$ denote
a tree, and $W$ denote a subset of elements in $\s$.  Given nodes
(elements) $a, b \in W$, we say that $b$ is \emph{consecutive} to node
$a$ w.r.t. $W$ in $s$ if the underlying poset is such that $a < b$ and
$a < w < b$ does not hold for any $w \in W$.  Furthermore, we use
$d_s(a, b)$ to denote the length of the path between $a$ and $b$ in
the Hasse diagram of the poset underlying $\s$.  The additional
auxiliary lemmas can now be stated as follows.
\begin{lemma}\label{lemma:bounding-the-height-without-W}
Let $\s$ be a $\Sigma$-tree. For every $m \in \mathbb{N}$, there
exists a subtree $t$ of $\s$ s.t. (i) the height of $t$ is at most $f(m,
|\Sigma|)$ and (ii) $t \equiv_m \s$.
\end{lemma}
\begin{lemma}\label{lemma:distance-reduction-between-root-and-a-node}
Let $s$ be a $\Sigma$-tree, $a$ be the root of $s$ and $b$ be any node
(element) of $s$.  For every $m \in \mathbb{N}$, there exists a
subtree $t$ of $s$ containing $a$ and $b$ such that (i) $d_t(a,
b) \leq f(m, 3 \cdot f(m, |\Sigma|))$ and (ii) $(t, b) \equiv_m (s,
b)$.
\begin{lemma}\label{lemma:bounding-distance-between-elements-of-W}
Let $s$ be a $\Sigma$-tree and $W$ be a subset of nodes (elements)
such that $|W| \le k$.  Let $a, b \in W$ be s.t. $b$ is consecutive to
$a$ w.r.t. $W$ in $s$. For every $m \in \mathbb{N}$, there exists
a subtree $t$ of $s$ s.t. (i) $t$ contains $W$ (ii) $t \equiv_m s$,
(iii) $b$ is consecutive to $a$ w.r.t. $W$ in $t$, and (iv) $d_t(a,
b) \leq (k - 1)\cdot (f(m, 3 \cdot f(m, |\Sigma|)) + 2)$.
\end{lemma}
\end{lemma}
Given these auxiliary lemmas,
Lemma~\ref{lemma:bounding-the-degree-and-height-given-W}(b) can be
proved as follows.  For notational clarity, define $N_1 = k \cdot
(f(m, 3 \cdot f(m, |\Sigma|)) + 2)$, $N_2 = k \cdot N_1$, $N_3 = f(m,
|\Sigma|)$ and $N_4 = N_2 + N_3 + 1$.  Let us further define $W_1
= \{root_s\} \cup W$.  By repeatedly applying
Lemma~\ref{lemma:bounding-distance-between-elements-of-W} with $W_1$
in place of $W$, we first obtain a subtree $z$ of $s$ s.t. (i) $z$
contains $W_1$, (ii) $z \equiv_m s$, (iii) $root_z = root_s$, and
(iii) $d_z(root_z, a) \le N_2$, for each element $a$ of $W$.  By
repeatedly applying Lemma~\ref{lemma:bounding-the-height-without-W} to
the subtrees rooted at elements $b$ s.t $d_z(root_z, b) = N_2+1$ and
using Lemma \ref{lemma:composition-lemma-for-trees}, we obtain a
subtree $t$ of $z$ s.t. (i) $t$ contains $W_1$ (ii) $t \equiv_m z$ and
(iii) $t$ has height at most $N_4$.  This proves
Lemma~\ref{lemma:bounding-the-degree-and-height-given-W}(b).

We now turn to proving the last three auxiliary lemmas referred to
above.  Lemma~\ref{lemma:bounding-the-height-without-W} is the easiest
to prove.  Let $A$ be the underlying set of $\s$. Define the function
$g: A \rightarrow$ $\Delta(m, \Sigma)$ as follows.  For $a \in A$,
$g(a)$ is the $\equiv_m$ class of $s_{\ge a}$. If no two elements in
any path in $\s$ have the same $g$ value, then the height of $\s$ is
at most $f(m, |\Sigma|)$, and the subtree $t$ required in
Lemma~\ref{lemma:bounding-the-height-without-W} can be chosen to be
$s$ itself. Otherwise, there exist $a, b \in A$ s.t. (i) $s \models
(a \leq b)$ and (ii) $g(a) = g(b)$.  If $a$ is the root of $\s$, then
let $s_1 = s_{\ge b}$ and we repeat the reasoning of this proof with
$s_1$ in place of $s$ (since $s_1 \equiv_m s$). Otherwise, let $c$ be
the parent of $a$ in $s$. By
Lemma \ref{lemma:composition-lemma-for-trees}, $s = \big(s_{\not\ge
a} \cdot_c s_{\ge a}\big) \equiv_m \big(s_{\not\ge a} \cdot_c s_{\ge
b}\big)$.  Let $s_1$ denote $\big(s_{\not\ge a} \cdot_c s_{\ge
b}\big)$.  Then $s_1$ is a subtree of $s$ and has fewer elements than
$s$.  We now repeat the same reasoning as above with $s_1$ in place of
$s$.  Since $s$ has only finitely many elements, this process
terminates with the desired subtree $t$ of $s$. \epf
\vspace{2pt}

The following is an easy corollary of
Lemma~\ref{lemma:bounding-the-height-without-W}.
\begin{corollary}\label{corollary:words-satisfy-Plogic(S, 0)}
Let $w$ be a given word over $\Sigma$. Then given $m \in \mathbb{N}$,
there exists a subword $v$ of $w$ s.t. (i) $|v| \leq f(m, |\Sigma|)$ and (ii)
$v \equiv_m w$.
\end{corollary}

Our proof of
Lemma~\ref{lemma:distance-reduction-between-root-and-a-node} is
inspired by the technique of companion models described in
~\cite{dawar-pres-under-ext}.  Recall that $a$ is the root of a
$\Sigma$-tree $s$ and $b$ is an arbitrary node in $s$ in
Lemma\ref{lemma:distance-reduction-between-root-and-a-node}.  For
notational clarity, let $q = f(m, 3 \cdot f(m, |\Sigma|))$.  If
$d_s(a, b) \leq q$, we choose $t = s$, and we are done. Otherwise,
suppose $d_s(a, b) = n > q$.  For $i \in \{0, n\}$, let $c_i$ be the
$i^{th}$ node along the (unique) path between $a$ and $b$.  Therefore,
$c_0 = a$, $c_n = b$ and $d_s(a, c_i) = i$.  Let the subtree rooted at
$c_i$ be denoted $t_i$, i.e. $t_i = s_{\ge c_i}$.  Let $z_0,
z_1, \ldots z_n$ be a sequence of subtrees of $s$ defined as follows:
(i) for $i \in \{0, \ldots n-1\}$, $z_i = (t_i)_{\not\ge c_{i+1}}$,
i.e. the subtree of $t_i$ obtained after removing the subtree
$t_{i+1}$, and (ii) $z_n = t_n$.  It is easy to see that $s$ can be
represented as $(((z_0 \cdot_a z_1) \cdot_{c_1}
z_2) \ldots \cdot_{c_{n-1}} z_n)$.  We now construct an $(n+1)$-length
word $w$ over the vocabulary $\Delta(m, \Sigma) \times \{0, 1, 2\}$ as
follows: (i) $w[1]$ = $(\delta, 1)$, where $\delta$ is the $\equiv_m$
class of $z_0$, (ii) for $i \in \{1, \ldots, n-1\}$, $w[i]$ =
$(\delta, 0)$, where $\delta$ is the $\equiv_m$ class of $z_i$, and
(iii) $w[n+1] = (\delta, 2)$, where $\delta$ is the $\equiv_m$ class
of $z_n$. By Corollary \ref{corollary:words-satisfy-Plogic(S, 0)},
there is a subword $w_1$ of $w$ s.t. (i) $|w_1| \leq q$, and (ii)
$w_1 \equiv_m w$.  Let the length of $w_1$ be $l+2$.  Observe that
$w_1$ must contain $w[1]$ and $w[n+1]$.  Since $w_1$ is a subword of
$w$, there exist positions $i_0, i_1, \ldots , i_l, i_{l+1}$ of $w$
s.t $i_0 = 1, i_{l+1} = n+1$ and $i_0 < \ldots < i_{l+1}$, and
$w_1[j+1] = w[i_{j}]$ for $j \in \{0, \ldots l+1\}$.  Now consider the
subtree $t$ defined by $(((z_0 \cdot_a z_{i_1}) \cdot_{c_{i_1}}
z_{i_2}) \ldots \cdot_{c_{i_l}} z_{n+1})$.  Observe that $t$ contains
$a$ and $b$ and $d_t(a, b) = |w_1| - 1 < q$. It is easy to see that
$(t, b) \equiv_m (s, b)$ (by similar arguments as
in~\cite{dawar-pres-under-ext}).\epf
\vspace{2pt}

Finally, Lemma~\ref{lemma:bounding-distance-between-elements-of-W} is
proved as follows.  Recall that $s$ is a $\Sigma$-tree, $W$ is subset
of at most $k$ nodes of $s$, and $a, b \in W$ are s.t.  $b$ is
consecutive to $a$ w.r.t $W$ in $s$.  For notational clarity, let $N_1
= f(m, 3 \cdot f(m, |\Sigma|))$ and $r = (k -1) \cdot (N_1 + 2)$.  If
$d_s(a, b) \leq r$, we choose $t = s$, and we are done.  Otherwise, we
show that there exists a subtree $s'$ of $s$ s.t. (i) $s'$ contains
$W$ (ii) $s' \equiv_m s$, (iii) $b$ is consecutive to $a$
w.r.t. $W$ in $s'$ and (iv) $d_{s'}(a, b) < d_s(a, b)$.
Lemma~\ref{lemma:bounding-distance-between-elements-of-W} then follows
by recursively applying the same reasoning to $s'$.

We reuse the notation $c_i$ and $z_i$ introduced in the proof of
Lemma \ref{lemma:distance-reduction-between-root-and-a-node}, but we
no longer require $a$ to be the root of $s$.  Let
$I \subseteq \{0, \ldots, n\}$ be the set of all indices $i$ such that
$z_i$ contains an element of $W$.  Clearly $0, n \in I$.  Since
$|W| \leq k$, we have $|I| \leq k$. If $|j - i| \leq (N_1 + 2)$ for
every pair of consecutive indices $i, j$ in $I$, then $d_s(a, b) \leq
(k-1) \cdot (N_1 +2) = r$.  However, this violates our assumption
$d_s(a, b) > r$.  Therefore, there exist consecutive indices $i',
j' \in I$ s.t. $i' < j'$ and $|j' - i'| > N_1 + 2$. Let $i = i' + 1$
and $j = j' - 1$.  Consider the subtree $z$ obtained by removing the
subtree $s_{\ge c_{j'}}$ from $s_{\ge c_i}$.  The following hold for
$z$: (i) $c_{i}$ is the root of $z$, (ii) $c_{j} \in z$, and (iii) $z$
does not contain any element of $W$.  Applying
Lemma \ref{lemma:distance-reduction-between-root-and-a-node} with $z$,
$c_{i}$ and $c_{j}$ as inputs, we know that there exists a subtree $y$
of $z$ containing $c_{i}$ and $c_{j}$ such that (i) $d_{y}(c_{i},
c_j) \leq N_1$ and (ii) $(y, c_j) \equiv_m (z, c_j)$.  It is also easy
to see that $s_{\ge c_i} = z \cdot_{c_j} s_{\ge c_{j'}}$.  Since $(y,
c_j) \equiv_m (z, c_j)$, by
Lemma \ref{lemma:composition-lemma-for-trees}, we have $s_{\ge
c_i} \equiv_m (y \cdot_{c_j} s_{\ge c_{j'}})$.  Let $y_1$ denote
$y \cdot_{c_j} s_{\ge c_{j'}}$ and let $y_2$ denote the subtree
obtained by removing $s_{\ge c_i}$ from $s$.  Applying
Lemma~\ref{lemma:composition-lemma-for-trees} again, we get
$\big(y_2 \cdot_{c_{i'}} s_{\ge
c_i} \big) \equiv_m \big(y_2 \cdot_{c_{i'}} y_1\big)$.  Note that $s =
y_2 \cdot_{c_{i'}} s_{\ge c_i}$. Then the subtree $y_2 \cdot_{c_{i'}}
y_1$ serves as the $s'$ required at the end of the previous paragraph.
\epf

\tbf{Remark}: Let $\mc{P}_{MSO}^{comp}(\cl{S}, k)$ be the property obtained
by replacing $\equiv_m$ in the definition of
$\mc{P}_{logic}^{comp}(\cl{S}, k)$ with $\equiv_m^{MSO}$, where
$\mf{A} \equiv_m^{MSO} \mf{B}$ denotes that $\mf{A}$ and $\mf{B}$
agree on all MSO sentences of total rank (i.e. first order and second
order quantifiers) $m$. The same ideas as in the proofs above show
that $\mc{P}_{MSO}^{comp}({\trees}(\Sigma), k)$ (and hence,
$\mc{P}_{MSO}^{comp}({\words}(\Sigma), k)$ hold for each
$k \in \mathbb{N}$.

\section{Generating new classes of structures}\label{section:constructing-new-classes}

We consider two natural ways of generating new classes of structures
from a base class $\cl{S}$ of structures.  The primary result of this
section is that classes generated by these techniques inherit the
$\mc{P}_{wqo}(\cl{S}, k)$ and $\mc{P}^{comp}_{logic}(\cl{S}, k)$
properties of the base classes.  For technical reasons, we assume in
this section that $\tau$ has only predicate symbols (i.e. no constant
symbols).
\subsection{Using unary/binary operations on structures}
 
We focus on disjoint union $(\sqcup)$, complementation $(!)$,
cartesian product $(\times)$ and tensor product $(\otimes)$ on
$\tau$-structures in a base class $\cl{S}$.  The definitions of
$\sqcup$ and $\times$ are standard (see ~\cite{makowsky} and Appendix 
\ref{appendix:new-classes}). The definitions of $!$ and $\otimes$
below are inspired by their definitions in the context of graphs. Let
$\mf{A}$ and $\mf{B}$ be $\tau$-structures.
\begin{itemize}
\item The \emph{complement} of $\mf{A}$, denoted $!\mf{A}$, is defined
  as follows: (i) $\mathsf{U}_{!\mf{A}} = \mathsf{U}_{\mf{A}}$, and
  (ii) for every $n$-ary predicate $R$ in $\tau$, for every $n$-tuple
  $(a_1, \ldots a_n) \in \mathsf{U}_{\mc{A}}^n$, $!\mf{A} \models
  R(a_1, \ldots a_n)$ iff $\mf{A} \not\models R(a_1, \ldots a_n)$.
\item The \emph{tensor product} of $\mf{A}$ and $\mf{B}$, denoted
  $\mf{A} \otimes \mf{B}$, is the structure ${\mf{C}}$ defined as
  follows: (i) $\mathsf{U}_{\mf{C}} = \mathsf{U}_{\mf{A}} \times
  \mathsf{U}_{\mf{B}}$, and (ii) for each $n$-ary predicate $R$ in $\tau$,
  for each $n$-tuple $\big((a_1, b_1), \ldots, (a_n, b_n)\big)$ of
  $\mathsf{U}_{\mf{C}}$, we have ${\mf{C}} \models R\big(((a_1, b_1),
  \ldots, (a_n, b_n))\big)$ iff $\mf{A} \models R(a_1, \ldots, a_n)$
  \emph{and} $\mf{B} \models R(b_1, \ldots, b_n)$.
\end{itemize}

We list below examples of classes of structures that can be generated
by applying the above operations repeatedly on simple classes of
structures.  In all these examples, colours are assumed to come
from a finite set of colours.
\begin{enumerate}
\item The class of coloured graphs, where the edge relation of each
  graph represents an equivalence relation.
\item The class of coloured co-graphs.
\item The class of $r$-dimensional \emph{grids} for every $r \in
  \mathbb{N}$, where a \emph{grid} is a tensor product of linear
  orders.
\end{enumerate}

Let $\mathsf{Op} = \{ \sqcup, !, \times, \otimes\}$.  The following
properties of operations in $\mathsf{Op}$ are used crucially in
subsequent proofs.  The properties are easy to prove, and the proofs
are omitted for lack of space.  Let $\circledast$ be a binary
operation in $ \mathsf{Op}$ and $m \in \mathbb{N}$.
\begin{enumerate}[P1)]
\item If $\mf{A}_1 \subseteq \mf{B}_1$ and $\mf{A}_2 \subseteq
  \mf{B}_2$, then (i) $!\mf{A}_1 \subseteq \,!\mf{B}_1$ and (ii)
  $(\mf{A}_1 \circledast \mf{A}_2) \subseteq (\mf{B}_1 \circledast
  \mf{B}_2)$.
\item If $\mf{A}_1 \equiv_m \mf{B}_1$ and $\mf{A}_2 \equiv_m
  \mf{B}_2$, then (i) $!\mf{A}_1 \equiv_m !\mf{B}_1$ and (ii)
  $(\mf{A}_1 \circledast \mf{A}_2) \equiv_m (\mf{B}_1 \circledast
  \mf{B}_2)$.
\end{enumerate}

Given a class $\cl{S}$, let $!\cl{S}$ denote the class $\{ !\mf{A}
\mid \mf{A} \in \cl{S}\}$. Given classes $\cl{S}_1$ and $\cl{S}_2$ and
a binary operation $\circledast \in \mathsf{Op}$, let $\cl{S}_1
\circledast \cl{S}_2$ denote the class $\{ \mf{A} \circledast \mf{B}
\mid \mf{A} \in \cl{S}_1, \mf{B} \in \cl{S}_2\}$.  Using the above
properties, we can now show the following. The proofs are deferred to
Appendix \ref{appendix:baby-operations}.

\begin{lemma}\label{lemma:baby-operations-preserve-pwqo-and-plogic}
Let $\cl{S}_1, \cl{S}_2$ be classes of structures. Let
$\circledast$ be a binary operation in $\mathsf{Op}$ and $k \in
\mathbb{N}$.
\begin{enumerate}
\item If $\mc{P}_{wqo}(\cl{S}_i, k)$ holds for $i \in \{1, 2\}$, then
  for $i \in \{1, 2\}$, each of $\mc{P}_{wqo}(!\cl{S}_i, k)$ and
  $\mc{P}_{wqo}(\cl{S}_1 \circledast \cl{S}_2, k)$ holds.
\item If $\mc{P}_{logic}^{comp}(\cl{S}_i, k)$ holds for $i \in \{1,
  2\}$, then for $i \in \{1, 2\}$, each of
  $\mc{P}_{logic}^{comp}(!\cl{S}_i, k)$ and
  $\mc{P}_{logic}^{comp}(\cl{S}_1 \circledast \cl{S}_2, k)$ holds.
\end{enumerate}
\end{lemma}

Given $O \subseteq \mathsf{Op}$, define an \emph{operation tree over
  $O$} to be a finite rooted tree\footnote{We think of a tree in the
  poset-theoretic sense, as in Section \ref{section:words-and-trees}.}
whose leaf nodes are unlabelled and non-leaf nodes are labelled with
elements of $O$. Furthermore, if the label of a non-leaf node is $op$,
the number of its successors equals the arity of $op$.  An operation
tree takes structures as ``inputs'' at its leaf nodes and produces an
``output'' structure at its root in the natural way.  We say that the
output structure is produced by ``applying'' the operation tree to its
inputs.

Given a class $\cl{S}$ of structures satisfying $\mc{P}_{wqo}(\cl{S},
k)$ (resp. $\mc{P}_{logic}^{comp}(\cl{S}, k)$), it follows from Lemma
\ref{lemma:baby-operations-preserve-pwqo-and-plogic} that the class
$\cl{S}'$ of structures obtained by applying any fixed operation tree
to the structures of $\cl{S}$, also satisfies $\mc{P}_{wqo}(\cl{S}',
k)$ (resp. $\mc{P}_{logic}^{comp}(\cl{S}', k)$).  The same holds if
$\cl{S}'$ is the union of the classes of structures obtained by
applying operation trees of height at most $h$ (for a fixed $h$) to
the structures in $\cl{S}$, where $h \in \mathbb{N}$. However, there
are interesting classes of structures that can be generated only by
applying operation trees of arbitrary heights.  For example, the class
of all co-graphs is produced from the class of single vertex graphs by
applying all operation trees over $\{\sqcup, !\}$.  What can we then
say about properties of such classes?  We present below a technique to
address this question.

Given a class $\cl{S}$ of structures and $O \subseteq \mathsf{Op}$, an
\emph{expression tree over $(\cl{S}, O)$} is an operation tree over
$O$ whose leaf nodes have been labelled with specific structures from
$\cl{S}$. If $s$ is an expression tree over $(\cl{S}, O)$, let
$\mf{C}_s$ denote the structure represented by $s$ upto isomorphism.
Given a node $a \in s$, we denote the subtree of $s$ rooted at $a$ as
$s_{a}$.  We denote by $Z_{\cl{S}, O}$ the class of all structures
defined by all possible expression trees over $(\cl{S}, O)$.
\begin{theorem}\label{theorem:pwqo-and-plogic-are-preserved-for-sqcup-and-bowtie-for-all-k}
Let $\cl{S}$ be a given class of structures and let $O = \{ \sqcup,
!\}$. For each $k \in \mathbb{N}$,
\begin{enumerate}
\item if $\mc{P}_{wqo}(\cl{S}, k)$ holds, so does $\mc{P}_{wqo}(Z_{\cl{S}, O}, k)$.
\item if $\mc{P}^{comp}_{logic}(\cl{S}, k)$ holds, so does
  $\mc{P}^{comp}_{logic}(Z_{\cl{S}, O}, k)$.
\end{enumerate}
\end{theorem}
\ti{Proof}: W.l.o.g., assume that $\cl{S} = !\cl{S}$. For otherwise,
we work with the union of $\cl{S}$ and $!\cl{S}$ since (i) $Z_{\cl{S}, O}
= Z_{\cl{S} \cup !\cl{S}, O}$ and (ii) $\mc{P}_{wqo}(\cl{S}, k)$
implies $\mc{P}_{wqo}(\cl{S} \cup !\cl{S}, k)$ (likewise for
$\mc{P}_{logic}^{comp}$).

As the first step, we introduce a new operation $\bowtie$ defined as
follows. Given structures $\mf{A}$ and $\mf{B}$, $\mf{A} \bowtie
\mf{B} ~=~$ $!((!\mf{A})\, \sqcup \, (!\mf{B}))$. The reader can
verify that $\bowtie$ enjoys properties P1 and P2 mentioned earlier.
Let $O_1 = \{\sqcup, \bowtie\}$. An additional important property,
call it P3, of $\circledast \in O_1$ is that $\mf{A} \subseteq \mf{A}
\circledast \mf{B}$ and $\mf{B} \subseteq \mf{A} \circledast \mf{B}$.
Given $\mf{A} \in Z_{\cl{S}, O}$, let $s'$ be an expression tree over
$(\cl{S}, O)$ s.t.  $\mf{C}_{s'} = \mf{A}$. Using the facts that $! (!
\mf{A}) = \mf{A}$ and $!  (\mf{A} \sqcup \mf{B}) = ((!\mf{A})\,
\bowtie \, (!\mf{B}))$, we can `push' the $!$ operator down to the
leaves to get an expression tree $s$ over $(\cl{S}, O_1)$
s.t. $\mf{C}_s = \mf{A}$.

To prove part (1) of the theorem, we use the notation $\cl{S}^k$
introduced in Lemma~\ref{lemma:towards-two-equivalent-defns-of-P-wqo}.
We show that if $Y = Z_{\cl{S}, O}$, then $Y^k$ is a w.q.o. under the
embedding relation.  Let $(\mf{A}_1, P_1), (\mf{A}_2, P_2), \ldots$ be
an infinite sequence of structures from $Y^k$, where each $P_i$ is an
atmost $k$-sized subset of $U_{\mf{A}_i}$.  Let $s_i$ be an expression
tree over $(\cl{S}, O_1)$ s.t. $\mf{C}_{s_i} = \mf{A}_i$.  Since only
$\sqcup$ and $\bowtie$ are used as labels of the non-leaf nodes in
$s_i$, every element of $P_i$ belongs to exactly one structure fed as
input at a leaf of $s_i$.  Let $t_i$ be the expression tree over
$(\cl{S}^k, O_1)$ s.t. $\mf{C}_{t_i} = (\mf{A}_i, P_i)$.  Let
$\hookrightarrow$ be the embedding relation over $\cl{S}^k$ and let
$\ll$ be the relation over $\cl{S}^k \cup O_1$ defined as $\ll \,=\,
\hookrightarrow \cup \{(\circledast, \circledast) | \,\circledast \in
O_1\}$. Since $\mc{P}_{wqo}(\cl{S}, k)$ holds, $\cl{S}^k \cup O_1$ is
a w.q.o. under $\ll$. Applying Kruskal's tree theorem to $(t_i)_{i \ge
  1}$, there exists $t_i, t_j$, where $i < j$ and a subtree $t_j'$ of
$t_j$ exists s.t. (i) the operation trees corresponding to $t_i$ and
$t_j'$ are identical and (ii) if $\mf{B}^i_1, \ldots \mf{B}^i_l$ are
the leaves of $t_i$ and $\mf{B}^j_1 \ldots \mf{B}^j_l$ are the
corresponding leaves of $t_j'$, then $\mf{B}^i_1 \hookrightarrow
\mf{B}^j_1, \ldots \mf{B}^i_l \hookrightarrow \mf{B}^j_l$. Using
properties P1 and P3 of $\sqcup$ and $\bowtie$, it is easy to see that
$M_{t_i}$ embeds into $M_{t_j'}$, and hence into $M_{t_j}$.

2) We now show that $\mc{P}_{logic}^{comp}(Z_{\cl{S}, O}, k)$ holds.
Let the vocabulary of structures be $\tau$ and let $m \in \mathbb{N}$
be given. Let $\Delta_m$ be the set of all the equivalence classes of
the $\equiv_m$ relation over the class of all $\tau$-structures and
let $f(m) = |\Delta_m|$.  Consider $\mf{A} \in Z_{\cl{S}, O}$ and
suppose that $s$ is an expression tree over $(\cl{S}, O_1)$
s.t. $\mf{C}_s = \mf{A}$.  Let $W$ be a set of $\leq k$ elements from
$\mf{A}$.  We observe that for $l \leq k$, the set $W$ identifies
structures $\mf{B}_{1}, \ldots, \mf{B}_{l}$ at the leaves of $s$, such
that all the elements of $W$ can be located inside these structures.
We organize our proof in two parts: (I) We first show that there
exists a sub-expression-tree $t$ of $s$, of height at most $h = k
\times f(m)$, s.t. (i) $\mf{B}_{j}$ is a leaf of $t$ for each $j \in
\{1, \ldots, l \}$, (ii) $\mf{C}_t \subseteq \mf{A}$ and (ii)
$\mf{C}_t \equiv_m \mf{A}$. (II) We create a tree $t_1$ from $t$ by
replacing the leaves of $t$ with bounded sized $m$-equivalent
substructures ensuring that $\mf{C}_{t_1}$ contains $W$. We then show
that $\mf{C}_{t_1}$ is the desired bounded sized $m$-equivalent
substructure of $\mf{A}$, to complete the proof.

(I) Our approach for height reduction is similar to the one used in
the proof of Lemma \ref{lemma:bounding-the-height-without-W}.
Let $A$ be the set of all nodes (internal + leaf) of $s$. Define the
function $g: A \rightarrow$ $\Delta_m \times \{1, \ldots, k\}$ as: for
$a \in A$, $g(a) = (\delta, i)$ where (i) $\delta$ is the $\equiv_m$
class of $\mf{C}_{s_a}$ where $s_a$ is the subtree of $s$ rooted at
$a$ (ii) the number of leaves in $s_a$ that contain any element of
$W$, is exactly $i$. Now if no two elements in any path in $s$ have
the same $g$ value, then the height of $s$ is at most $h$, whence $t$
can be taken to be $s$ itself. Else there exist elements $a, b \in A$
s.t.  $b \in s_a$ and $g(a) = g(b)$. If $a$ is the root of $s$, then
let $s_1 = s_b$. Else, let $c$ be the parent of $a$ in $s$ and let
$s_1$ be the subtree of $s$ obtained by deleting $s_a$ and joining
$s_b$ to $c$ (i.e. by making $c$ the parent of $b$). In either case,
the following are true about $s_1$: (i) All of the $\mf{B}_{j}$s are
present as leaves of $s_1$ --- since $g(a) = g(b)$, it means that all
the leaves of $s_a$ that contain elements of $W$ are also leaves of
$s_b$. (ii) $\mf{C}_{s} \equiv_m \mf{C}_{s_1}$ --- If $s_1 = s_b$
above, then $\mf{C}_s = \mf{C}_{s_a} \equiv_m \mf{C}_{s_b} =
\mf{C}_{s_1}$. Else since the tree obtained by deleting $s_a$ from $s$
is the same as the tree obtained by deleting $s_b$ from $s_1$, we have
by the property P2 of $\sqcup$ and $\bowtie$ that $\mf{C}_{s} \equiv_m
\mf{C}_{s_1}$ (iii) $\mf{C}_{s_1} \subseteq \mf{C}_s$ -- this is
because $\mf{C}_{s_b} \subseteq \mf{C}_{s_a}$ due to property P3
mentioned above and because $\sqcup$ and $\bowtie$ have property P1.
We now repeat all of the argument above with $s_1$ in place of $s$. It
is clear that continuing this way, we get the desired subtree $t$ of
$s$.

(II) Let $\alpha$ be the computable function witnessing
$\mc{P}^{comp}_{logic}(\cl{S}, k)$. Since each leaf $\mf{B}$ of $t$ is
also a leaf of $s$, we have $\mf{B} \in \cl{S}$. Then for each leaf
$\mf{B}$, by the definition of $\mc{P}^{comp}_{logic}(\cl{S}, k)$,
there exists a substructure $\mf{B}^1$ of $\mf{B}$ of size $\leq
\alpha(m)$, s.t. $\mf{B}^1 \in \cl{S}$, $\mf{B}^1 \equiv_m \mf{B}$ and
all the elements of $W$ contained in $\mf{B}$ are also contained in
$\mf{B}^1$.  Let $t_1$ be the tree obtained by replacing each leaf
$\mf{B}$ of $t$ with $\mf{B}^1$. Then $t_1$ has the following
properties: (i) $\mf{C}_{t_1}$ contains $W$ (ii) $|\mf{C}_{t_1}| \leq
2^h \times \alpha(m)$ (iii) $\mf{C}_{t_1} \subseteq \mf{C}_t$ (by
property P1) (iv) $\mf{C}_{t_1} \equiv_m \mf{C}_t$ (by property P2)
and (v) $\mf{C}_{t_1} \in Z_{\cl{S}, O}$, because by expressing the
$\bowtie$ operator in terms of $\sqcup$ and $!$ as in its definition,
we get an expression tree $t'$ over $(\cl{S}, O)$ s.t. $\mf{C}_{t'} =
\mf{C}_{t_1}$.

Combining (I) and (II) above, we get that
$\mc{P}^{comp}_{logic}(Z_{\cl{S}, O}, k)$ holds, where the witnessing
computable function is $\beta(m) = 2^{(k \cdot f(m))} \times
\alpha(m)$.\epf

We do not know whether Theorem
\ref{theorem:pwqo-and-plogic-are-preserved-for-sqcup-and-bowtie-for-all-k}
holds for all $k$ if we consider $O = \mathsf{Op}$ in its
statement. However, for $k = 0$ or $1$, this result does go
through. We skip the proof in this paper.

\subsection{Constructing words and trees over classes of structures}

Given a class $\cl{S}$ of $\tau$-structures, we now define the classes
${\words}(\cl{S})$, resp. ${\trees}(\cl{S})$, which intuitively
speaking, are the classes of words, resp. trees, of structures in
$\cl{S}$.

Formally, a \emph{word over $\cl{S}$}, or simply, a $\cl{S}$-word, is
a finite sequence of structures from $\cl{S}$.  A $\cl{S}$-word $w =
\mf{A}_1 \cdot \mf{A}_2 \cdots \mf{A}_n$ has a natural representation
as a $\nu$-structure $\mf{B}$, where $\nu = \{\leq\} \cup \tau$: (i)
$\mathsf{U}_\mf{B} = \bigcup_{i =1}^{i = n} \mathsf{U}_{\mf{A}_i}$
(ii) the $\tau$-reduct of $\mf{B}(\mathsf{U}_{\mf{A}_i})$ is exactly
$\mf{A}_i$ for $i \in \{1, \ldots, n\}$ (iii) for each predicate $R
\in \tau$, of arity $k$, if $\bar{a}$ is a $k$-tuple from $U_\mf{B}$
having at least two components from two different $U_{\mf{A}_i}$s,
then $\mf{B} \models \neg R(\bar{a})$ (iv) $\mf{B} \models (a \leq b)$
for all $a \in \mathsf{U}_{\mf{A}_i}$ and $b \in
\mathsf{U}_{\mf{A}_j}$ for $1\leq i \leq j \leq n$ and $\mf{B} \models
\neg (b \leq a)$ for all $a \in \mathsf{U}_{\mf{A}_i}$ and $b \in
\mathsf{U}_{\mf{A}_j}$ for $1\leq i < j \leq n$. Then
${\words}(\cl{S})$ is just the class of all $\cl{S}$-words. The formal
definition of ${\trees}(\cl{S})$ is similar.

Given two $\cl{S}$-words $w_1$ and $w_2$ s.t. $w_1 = \mf{A}_1 \cdot
\mf{A}_2 \cdots \mf{A}_r$, it is easy to check given the definitions
above, that $w_1$ is embeddable in $w_2$ iff there exists a
sub-$\cl{S}$-word $\mf{B}_1 \cdot \mf{B}_2 \cdots \mf{B}_r$ of $w_2$
such that $\mf{A}_i$ embeds into $\mf{B}_i$ for each $i \in \{1,
\ldots, r\}$.  We now show the following.

\begin{prop}\label{prop:pwqo(S, k)-holds-for-words-and-trees-over-classes}
Let $\cl{S}$ be a given class of structures. Then given $k \in
\mathbb{N}$, the following are true.
\begin{enumerate}
\item $\mc{P}_{wqo}(\cl{S}, k) \rightarrow
  \mc{P}_{wqo}({\words}(\cl{S}), k)$.
\item $\mc{P}_{wqo}(\cl{S}, k) \rightarrow
  \mc{P}_{wqo}({\trees}(\cl{S}), k)$.
\end{enumerate}
\end{prop}

\ti{Proof}: We firstly observe that each of the above statements for
$k = 0$ follows straightaway from Higman's lemma and Kruskal's tree
theorem.  For $k > 0$, we show the result for ${\words}(\cl{S})$. The
proof for ${\trees}(\cl{S})$ is similar. Let $Y = {\words}(\cl{S})$.
Consider an infinite sequence $(\mf{A}_i, P_i)_{i \ge 1}$ from
$Y^k$. For each $i$, the elements of $P_i$ can be located within the
structures forming $\mf{A}_i$. Specifically, if $\mf{A}_i = \mf{A}_i^1
\cdots \mf{A}_i^n$, then for $j \in \{1, \ldots, n\}$, there exist
subsets $P_i^j$ of elements of $\mf{A}_i^j$ s.t. $(\mf{A}_i, P_i) =
(\mf{A}_i^1, P_i^1) \cdots (\mf{A}_i^n, P_i^n) \in
{\words}(\cl{S}^k)$. Since $\mc{P}_{wqo}(\cl{S}^k, 0)$, we have
$\mc{P}_{wqo}({\words}(\cl{S}^k), 0)$. Whence there exist $i, j$ where
$i < j$ s.t. if $(\mf{A}_i, P_i) = (\mf{A}_i^1, P_i^1) \cdots
(\mf{A}_i^n, P_i^n)$, then there is a sub-$\cl{S}^k$-word
$(\mf{B}_j^1, P_j^1) \cdots (\mf{B}_j^n, P_j^n)$ of $(\mf{A}_j, P_j)$
such that $(\mf{A}_i^l, P_i^l)$ embeds into $(\mf{B}_j^l, P_j^l)$ for
each $l \in \{1, \ldots, n\}$. Then $(\mf{A}_i, P_i)$ embeds into
$(\mf{A}_j, P_j)$.\epf

\begin{prop}\label{prop:plogic(S, k)-holds-for-words-and-trees-over-classes}
Let $\cl{S}$ be a given class of structures. Then given $k \in
\mathbb{N}$, the following are true.
\begin{enumerate}
\item $\mc{P}_{logic}^{comp}(\cl{S}, k) \rightarrow
  \mc{P}_{logic}^{comp}({\words}(\cl{S}), k)$.
\item $\mc{P}_{logic}^{comp}(\cl{S}, k) \rightarrow
  \mc{P}_{logic}^{comp}({\trees}(\cl{S}), k)$.
\end{enumerate}
\end{prop}

\ti{Proof}: We show the result for ${\words}(\cl{S})$.  The proof for
${\trees}(\cl{S})$ is similar. Let $m \in \mathbb{N}$ be
given. Consider $\mf{A} \in {\words}(\cl{S})$ s.t. $\mf{A} = \mf{A}_1
\cdots \mf{A}_r$.  Let $W$ be a set of at most $k$ elements of
$\mf{A}$. Let $W_i$ be the set of elements of $W$ that are contained
in $\mf{A}_i$ for $i \in \{1, \ldots, r\}$.  Let $\alpha$ be the
computable function witnessing $\mc{P}_{logic}^{comp}(\cl{S}, k)$.
For $i \in \{1, \ldots, r\}$, since $\mf{A}_i \in \cl{S}$, there
exists $\mf{B}_i \subseteq \mf{A}_i$ s.t. (i) $\mf{B}_i \in \cl{S}$
(ii) $\mf{B}_i$ contains $W_i$ (ii) $|\mf{B}_i| \leq \alpha(m)$ (iii)
$\mf{B}_i \equiv_m \mf{A}_i$. Then consider the $\cl{S}$-word $\mf{B}
= \mf{B}_1 \cdots \mf{B}_r$. It is easy to see that (i) $\mf{B} \in
{\words}(\cl{S})$ (ii) $\mf{B}$ contains $W$ (iii) $\mf{B} \subseteq
\mf{A}$ and (iv) $\mf{B} \equiv_m \mf{A}$.

Since each $\mf{B}_i$ has size at most $\alpha(m)$, $\mf{B}$ can be
treated as a word $u$ over a finite alphabet $\Sigma_m$, where
$\Sigma_m$ is the set of all $\tau$-structures (upto isomorphism) of
size at most $\alpha(m)$. Note that there exists a computable function
$\beta:\mathbb{N} \rightarrow \mathbb{N}$ s.t.  $|\Sigma_m| =
\beta(m)$.  Let $W^1 = \{ i \mid \mf{B}_i ~\text{contains at least one
  element of}~W_i\}$. Clearly $|W^1| \leq k$.  Then by Proposition
\ref{prop:words-and-trees-satisfy-plogic(S, k)}, there exists a
subword $v$ of $u$ containing $W^1$ such that $v \equiv_m u$ and $|v|
\leq \gamma(m, |\Sigma_m|)$ for some computable function $\gamma$.  If
$v = u\left[i_1\right] \cdots u\left[i_l\right]$, then consider the
$\cl{S}$-word $\mf{B}^1 = \mf{B}_{i_1} \cdots \mf{B}_{i_l}$. It is
easy to show that $\mf{B}^1 \equiv_m \mf{B}$.  Further,
we see that (i) $\mf{B}^1$ contains $W$ (ii) $\mf{B}^1 \subseteq
\mf{B}$ (iii) $\mf{B}^1 \in {\words}(\cl{S})$ (iv) $|\mf{B}^1| \leq
\alpha(m) \times \gamma(m, \beta(m))$. \epf

Two examples of classes of structures that can be generated using the
above operations are the class of all coloured total pre-orders and
the class of all coloured pre-order trees.

\section{Conclusion}\label{sec:conclusion}
We studied two abstract properties of classes of finite structures
that allow a generalization of the classical {\lt} theorem to hold.
This augments earlier work on identifying properties of classes of
finite structures that allow the {\lt} theorem to hold.  We showed
that several interesting classes of finite structures satisfy the
properties discussed in this paper, and even allow an effective
translation of a sentence in $PSC(k)$ to an equivalent
$\exists^k\forall^*$ sentence.  Nevertheless, several questions remain
open.  For example, while we have shown that $\mc{P}_{wqo}(\cl{S}, 0)$
implies ${\PSC}(k) = \exists^k\forall^*$, we do not know yet whether
the implication is strict.  Similarly, all properties studied in this
paper lead to cores being definable by a relativized formula.  We
would like to study more closely the relationship between $PSC(k) =
\exists^k\forall^*$ and definability of cores, not necessarily by
relativized formulae.

\section*{Acknowledgment}
The authors thank Ajit A. Diwan for insightful
discussions.

\bibliographystyle{plain}
\bibliography{refs} 
\begin{appendix}
\section{Proof of Lemma \ref{characterization-of-PSC(0)}}\label{appendix:char-of-PSC(0)}

All the arguments below are over $\cl{S}$.

Suppose $\mc{P}_{wqo}(\cl{S}, 0)$ holds. Let $\cl{C} \in {\PS}$. Then
the complement $\overline{\cl{C}}$ of $\cl{C}$ is preserved under
extensions. Let $\cl{H}$ be the class of minimal models of
$\overline{\cl{C}}$. If $\cl{H}$ is finite upto isomorphism, then
taking the disjunction of the existential closures of the `atomic
diagrams' of the models of $\cl{H}$, we get a $\exists^*$ sentence
$\psi$ defining $\overline{\cl{C}}$. Then $\neg \psi$ defines
$\cl{C}$. We show that $\cl{H}$ must always be finite. For if not,
then let $(\mf{A}_i)_{i \ge 1}$ be an infinite sequence of structures
from $\cl{H}$ where the $\mf{A}_i$ are all distinct. Since
$\mc{P}_{wqo}(\cl{S}, 0)$ holds, we have for some $i, j$ s.t. $i < j$,
that $\mf{A}_i \hookrightarrow \mf{A}_j$. But this violates the
minimality of $\mf{A}_j$.

Suppose ${\PS} \equiv \forall^*$. Let $(\mf{A}_i)_{i \ge 1}$ be an
infinite sequence of structures from $\cl{S}$. Let $\cl{C}
= \{\mf{B} \in \cl{S} \mid
\mf{A}_i \hookrightarrow \mf{B} ~\text{for some}~i \ge 1\}$. Then $\cl{C}$
is preserved under extensions, whence $\overline{\cl{C}} \in
{\PS}$. By assumption then, there exists a $\forall^*$ sentence $\psi$
defining $\overline{\cl{C}}$. Then $\cl{C}$ is defined by an
$\exists^*$ sentence, namely $\neg \psi$. Then the number of minimal
models of $\cl{C}$ is finite. But the minimal models of $\cl{C}$ are
exactly the $\mf{A}_i$s. Clearly then for some $i, j$ s.t. $i < j$, we
have $\mf{A}_i \cong
\mf{A}_j$. \epf\\

\section{Witness function for $\mc{P}_{logic}(\cl{S}, k)$ lies in the second level of the Turing heirarchy}\label{appendix:turing-hierarchy}
To see why this is so, note that we can construct a Turing machine
$TM_{\cl{S}}$ that accepts $m$, $k$, $p$ as inputs, recursively
enumerates $S$ and halts only if it finds a structure $\mf{A}$ in
$\cl{S}$ and a subset $W$ of $U_{\mf{A}}$ of size $\le k$, such that
no substructure $\mf{B}$ of $\mf{A}$, sized $p$ or less, satisfies the
conditions of Definition~\ref{defn:Plogic(S,k)}.  Using an oracle that
answers whether $TM_{\cl{S}}(m, k, p)$ halts, we can now construct a
Turing machine that accepts $m, k$ as inputs, enumerates values of
$p$, and outputs the first $p$ for which $TM_{\cl{S}}(m, k, p)$
doesn't halt.

\section{Proof of Proposition \ref{prop:p-comp-logic-does-not-imply-pwqo}}\label{appendix:cycles+paths}

Let $C_n$, resp. $P_n$, denote an undirected cycle, resp. path, of
length $n$. Let $mP_n$ denote the disjoint union of $m$ copies of
$P_n$. Let $H_n = \bigsqcup_{i = 0}^{i = 3^n} nP_i$ and $G_n = C_{3^n}
\sqcup H_n$, where $\bigsqcup$ and $\sqcup$ denote disjoint union.
Then consider the class $\cl{S}$ of undirected graphs given by $\cl{S}
= \cl{S}_1 \cup \cl{S}_2$, where $\cl{S}_1 = \{H_n \mid n \ge 1\}$ and
$\cl{S}_2 = \{ G_n \mid n \ge 1\}$.  We will show that $\cl{S} \in
\Gamma^*_{comp} \setminus \Gamma^0_{wqo}$.

The fact that $\cl{S}$ is not in $\Gamma^0_{wqo}$ is easy to see.
Consider the sequence $(G_n)_{n \ge 1}$. The cycles in these graphs
prevent any $G_m$ from being embeddable in any $G_n$ for $m \neq n$.

We now show $\cl{S} \in \Gamma^*_{comp}$ by showing that given $k \in
\mathbb{N}$, the function $f:\mathbb{N} \rightarrow \mathbb{N}$
defined as $f(m) = |G_{m + k + 2}|$ witnesses
$\mc{P}^{comp}_{logic}(\cl{S}, k)$.  Towards this, we will first need
some basic facts about $C_n$, $P_n$, $H_n$ and $G_n$. Let $m \in
\mathbb{N}$ be given.
\begin{enumerate}
\item If $n_1, n_2 \ge 3^m$, then $P_{n_1} \equiv_m P_{n_2}$.
\item If $n_1 \ge 3^m$ and $n_2 \ge m$, then $n_2P_{n_1} \equiv_m
  mP_{3^m}$.
\item If $n_1 \leq n_2$, then $H_{n_1}$ always embeds in $H_{n_2}$.
\item If $m \leq n_1 \leq n_2$, then $H_{n_1} \equiv_m
  H_{n_2}$. (follows from (1) and (2) above)
\item If $n \ge m$, then $G_n \equiv_m H_n$.
\end{enumerate}

We will need the following two helper lemmas.
\vspace{2pt}

\begin{lemma}\label{lemma:bounded-subst-of-a-path-containing-W}
Given $m, k \in \mathbb{N}$, a path $P$ of length $\ge 3^{m + k + 2}$
and a set $W$ of $l \leq k$ nodes of $P$, there exists a substructure
$G$ of $P$ containing $W$, such that $G$ is a disjoint union of at
most $l$ paths, each path having length at most $3^{m + k + 2}$.
\end{lemma}
\emph{Proof}: If $W = \{a\}$ for some node $a$, then take $G$ to be
the substructure induced by $a$. Else, let $W = \{a_1, \ldots, a_l\}$
be the given set of $l$ nodes of $P$ where $1 < l \leq k$.  Let $a_0$
and $a_{l+1}$ be the end points of $P$. W.l.o.g., assume that for $j
\in \{0, l\}$, $a_j$ and $a_{j+1}$ are consecutive, i.e. there is no
$a_{j_1}$ that is strictly `in between' $a_j$ and $a_{j+1}$ for $j_1
\in \{0, l\}$.

If the distance between $a_j$ and $a_{j+1}$ is at most $3^{m + 1}$ for
any $j$, then the distance between $a_1$ and $a_l$ is at most $(k - 1)
\times 3^{m + 1} \leq 3^{m + k + 2}$. Then taking $G$ to be the
substructure induced by all the points of $P$ that lie in between and
include $a_1$ and $a_l$, we see that $G$ is indeed as desired. Else
for some $j$, the distance between $a_j$ and $a_{j+1}$ is $> 3^{m +
1}$. Then let $b_1$, resp. $b_2$, be the point lying between $a_j$ and
$a_{j+1}$ at a distance of exactly $3^m$ from $a_j$,
resp. $a_{j+1}$. Clearly $b_1$ and $b_2$ cannot be adjacent. Then let
$P_1$ be the path obtained by taking the substructure of $P$ induced
by all the points lying between and including $a_0$ and
$b_1$. Likewise let $P_2$ be the path obtained by taking the
substructure of $P$ induced by all the points lying between and
including $b_2$ and $a_{l+1}$. Then $P_1$ contains $j$ points of $W$
and $P_2$ contains $l - j$ points of $W$. Then we can perfom the above
reasoning recursively on $P_1$ and $P_2$.  Let $G_1$ and $G_2$
respectively be the substructures of $P$ obtained by doing the
aforesaid reasonings on $P_1$ and $P_2$. Then $G_1$, resp. $G_2$,
contains $a_1, \ldots, a_j$, resp. $a_{j+1}, \ldots, a_l$, and is a
disjoint union of at most $j$ paths, resp. at most $l - j$ paths, each
path having length at most $3^{m + k + 2}$. Then check that $G = G_1
\sqcup G_2$ is indeed as desired.\epf
\vspace{2pt}

\begin{lemma}\label{lemma:bounded-subst-of-a-cycle-containing-W}
Given $m, k \in \mathbb{N}$, a cycle $C$ of length $\ge 3^{m + k + 2}
+ 2$ and a set $W$ of $l \leq k$ nodes of $C$, there exists a
substructure $G$ of it containing $W$, such that $G$ is a disjoint
union of at most $l$ paths, each of length at most $3^{m + k + 2}$.
\end{lemma}
\emph{Proof}: Clearly some node of $C$ is not in $W$. Deleting this
node, we get a path of length $\ge 3^{m + k + 2}$ that contains
$W$. Invoking Lemma
\ref{lemma:bounded-subst-of-a-path-containing-W}, we are done. \epf

We are now ready to prove Proposition
\ref{prop:p-comp-logic-does-not-imply-pwqo}. Consider a structure $\mf{A}
\in \cl{S}$ and let $W$ be a set of at most $k$ elements of $\mf{A}$. We
have two cases: (a) $\mf{A} \in \cl{S}_1$ (b) $\mf{A} \in \cl{S}_2$.
\vspace{2pt}

\underline{$\mf{A} \in \cl{S}_1$:}

Then $\mf{A} = H_n$ for some $n$. If $n \leq (m + k + 2)$, then taking
$\mf{B}$ to be $\mf{A}$, we see that the conditions of
$\mc{P}^{comp}_{logic}(\cl{S}, k)$ are satisfied for $\mf{A}$. Else $n
> (m + k + 2) $. Let $W^1$ be the subset of $W$ contained in the paths
of $\mf{A}$, of lengths $\leq 3^{m + k + 2}$, and let $W^2$ be the
subset of $W$ contained in paths of $\mf{A}$, of lengths $>
3^{m+k+2}$.

If $W^2 = \emptyset$, then all of $W$ is contained within the paths of
lengths $\leq 3^{m + k + 2}$. Since $n > (m + k + 2)$, it is easy to
see that taking $m + k + 2$ paths of each length from 0 to $3^{m + k +
2}$ such that $W$ is contained in these paths, we get a substructure
$\mf{B}$ of $\mf{A}$ that contains $W$ and that is isomorphic to $H_{m
+ k + 2}$.

If $W^2 \neq \emptyset$, then let $P_1, \ldots, P_l$ be the $l \leq k$
paths of lengths $> 3^{m+ k +2}$ that contain some element of $W^2$.
Applying Lemma \ref{lemma:bounded-subst-of-a-path-containing-W} to
each $P_i$ for $i \in \{1, \ldots, l\}$ and taking the disjoint union
of the substructures $G_i$ of these obtained thereof, we get a
substructure $G^1$ of $G$ containing $W^2$ s.t. $G^1$ is a disjoint
union of at most $k$ paths of lengths at most $3^{m + k + 2}$. Now as
in the previous case, we can construct a substructure $N_1$ of
$\mf{A}$ containing $W^1$ such that $\mf{B}_1$ is isomorphic to $H_{m
+ k + 2}$.  Since $\mf{B}_1$ contains $m + k + 2$ paths of each length
$i$ ranging from 0 to $3^{m + k + 2}$, it is easy to see that
`swapping' the paths in $G^1$ with paths of same lengths in $\mf{B}_1$
that do not contain any element of $W^1$, we get a substructure
$\mf{B}$ of $\mf{A}$ containing all of $W$ and which is isomorphic to
$H_{m + k + 2}$.

In either case, we get a substructure $\mf{B}$ of $\mf{A}$ that
contains $W$ and is isomorphic to $H_{m + k + 2}$.  Then
$\mf{B} \in \cl{S}$.  By the observations at the outset,
$\mf{B} \equiv_m \mf{A}$. Further, it is clear that $|\mf{B}| \leq
f(m)$. Then $\mf{A}$ satisfies the conditions of
$\mc{P}^{comp}_{logic}(\cl{S}, k)$.\vspace{2pt}

\underline{$\mf{A} \in \cl{S}_2$}:

Then $\mf{A} = C_{3^n} \sqcup H_n$. If $n \leq (m+ k + 2)$, then
taking $\mf{B}$ to be $\mf{A}$, we see that the conditions of
$\mc{P}^{comp}_{logic}(\cl{S}, k)$ are satisfied for $\mf{A}$. Else $n
> (m + k + 2) $. Let $W^1$ be the subset of $W$ contained in $C_{3^n}$
and let $W^2$ be the subset of $W$ contained in $H_n$. By the same
argument as in the previous case, we can show that there exists a
substructure $\mf{B}_1$ of $H_n$ containing $W^2$ such that $\mf{B}_1$
is isomorphic to $H_{m + k + 2}$. Applying Lemma
\ref{lemma:bounded-subst-of-a-cycle-containing-W} to $C_{3^n}$, we get
a substructure $G^1$ of $C_{3^n}$ that contains $W^1$ and such that
$G^1$ is a disjoint union of at most $|W^1|$ paths, each of length at
most $3^{m + k + 2}$. Since we have $m + k + 2$ paths of each length
$i$ ranging from 0 to $3^{m + k + 2}$ in $\mf{B}_1$, it is easy to see
that `swapping' the paths in $G^1$ with paths of same lengths in
$\mf{B}_1$ that do not contain any element of $W^2$, we get a
substructure $\mf{B}$ of $\mf{A}$ containing all of $W$ and that is
isomorphic to $H_{m + k + 2}$.  Then $\mf{B} \in \cl{S}$.  By the
observations above $\mf{B} \equiv_m \mf{A}$. Further, it is clear that
$|\mf{B}| \leq f(m)$. Then $\mf{A}$ satisfies the conditions of
$\mc{P}^{comp}_{logic}(\cl{S}, k)$.\epf

\section{Proofs of relations between various classes}\label{appendix:relations}

That $\cl{S}_4 \in \Gamma^*_{comp} \setminus \Gamma^0_{wqo}$ and
$\cl{S}_5 \in \Gamma^*_{wqo} \setminus \Gamma^0_{comp}$ follow from
Propositions \ref{prop:p-comp-logic-does-not-imply-pwqo}
and \ref{prop:pwqo-does-not-imply-p-comp-logic}. We show below the
results conceerning $\cl{S}1$ and $\cl{S}_3$.

\underline{$\cl{S}_1 \in \Gamma^0_{wqo}$, $\cl{S}_3 \in \Gamma^0_{wqo}$}

We show the reasoning for $\cl{S}_3$. The result for $\cl{S}_1$ would
follow.

Let $(G_i)_{i \ge 1}$ be an infinite sequence of graphs from
$\cl{S}_3$.  Consider $G_1$ and let the size of $G_1$ be $n$. If for
some $i > 1$, $G_i$ contains a path of length $2n$, then it is easy to
see that $G_1 \longmapsto G_i$. Else, for all $i > 1$, all paths in
$G_i$ are of length less than $2n$. Then construct a $2n$-tuple $s_i$
corresponding to $G_i$, where the $j^{\text{th}}$ component of $s_i$
is the number of paths of length $j$ in $G_i$, for $0 \leq j <
2n$. Now applying Dickson's lemma to the sequence $(s_i)_{i > 1}$, we
get that $s_i$ is component-wise less than $s_j$ for some $i < j$, $i
> 1$. Clearly then $G_i \hookrightarrow G_j$. This shows that
$\cl{S}_3$ belongs to $\Gamma_{wqo}^0$.

\underline{$\cl{S}_1 \in \Gamma^0_{comp}$, $\cl{S}_3 \in \Gamma^0_{comp}$}

From the facts about paths mentioned in Appendix 
\ref{appendix:cycles+paths}, it follows that:
\begin{enumerate}
\item The computable function $f: \mathbb{N} \rightarrow \mathbb{N}$ given
by $f(m) = 3^m$ witnesses $\cl{S}_1 \in \Gamma^0_{comp}$.
\item The computable function $f: \mathbb{N} \rightarrow \mathbb{N}$ given
by $f(m) = m \times (1 + 2 + \ldots + 3^m)$ witnesses
$\cl{S}_1 \in \Gamma^0_{comp}$.
\end{enumerate}

\underline{$\cl{S}_1 \in \Gamma^* \setminus \Gamma^*_{logic}$, $\cl{S}_3 \notin \Gamma^*$}

That $\cl{S}_3 \notin \Gamma^*$ was already shown in
~\cite{wollic12-paper}. That $\cl{S}_1 \in \Gamma^*$ can be shown by a
reasoning similar to that in the proof of
Lemma \ref{lemma:cycles}. That $\cl{S}_1 \notin \Gamma^*_{logic}$ can
be seen as follows. Towards a contradiction, suppose
$\cl{S}_1 \in \Gamma^*_{logic}$. Then for each $k$, there exists a
function $f_k$ witnessing $\mc{P}_{logic}(\cl{S}_1, k)$. Let $k = 2$
and consider $f_2$. Given $m$, consider $\mf{A} \in \cl{S}_1$
s.t. $\mf{A}$ is a path of length $> f_2(m)$. Let $W$ be the set
consisting exactly of the end points of $\mf{A}$. It is clear that the
only substructure of $\mf{A}$ that contains $W$ and that is in
$\cl{S}_1$ is $\mf{A}$ itself. Then $\mf{A}$ does not contain any
substructure of size $\leq f_2(m)$ which is $m$-equivalent to
$\mf{A}$. This contradicts the assumption that $f_2$ witnesses
$\mc{P}_{logic}(\cl{S}_1, 2)$.

\section{Proof of Lemma \ref{lemma:towards-two-equivalent-defns-of-P-wqo}}\label{appendix:alt-defn-of-Pwqo}

We will show below that $\mc{P}_{wqo}(\cl{S}^k, 0)$ implies
$\mc{P}_{wqo}(\cl{S}_k, 0)$. We say that an infinite sequence $I_1$
from $\cl{S}_k$ is `good' if there exist $i_1, i_2, i_3, \ldots$ where
$i_1 < i_2 < i_3 < \ldots$ s.t. the structure at index $i_1$ in $I_1$
embeds in the structure at index $i_2$ in $I_1$, which in turn embeds
in the structure at index $i_3$ in $I_1$, and so on. We will show that
every infinite sequence from $\cl{S}_k$ is good.

We will first prove two helper lemmas.

\begin{lemma}\label{lemma:expansion-with-permutations}
Let $I = (\mf{A}_i, \bar{a}_i)_{i \ge 1}$ be an infinite sequence from
$\cl{S}_k$ such that for any $i$, the components of $\bar{a}_i$ are
distinct. Let $\Pi_1, \ldots, \Pi_{k!}$ be permutations of the set
$\{1, \ldots, k\}$ and let $J$ be the sequence given by $J =
(\mf{A}_1, \Pi_1(\bar{a}_1)), \ldots,
(\mf{A}_1, \Pi_{k!}(\bar{a}_1)),$
$(\mf{A}_2, \Pi_1(\bar{a}_2)), \ldots,$
$(\mf{A}_2, \Pi_{k!}(\bar{a}_2)), \ldots$. Then $I$ is good iff $J$ is
good.
\end{lemma}
\emph{Proof}: If $I$ is good, then clearly $J$ is good. Suppose $J$ 
is good. Then there must exist some
$\Pi \in \{\Pi_1, \ldots, \Pi_{k!}\}$ and some infinite subsequence
$(\mf{A}_{i_r}, \Pi(\bar{a}_{i_r}))_{r \ge 1}$
s.t. $(\mf{A}_{i_p}, \Pi(\bar{a}_{i_p}))$ embeds in
$(\mf{A}_{i_r}, \Pi(\bar{a}_{i_r}))$ each $p < r$, $p \ge 1$. Then
$(\mf{A}_{i_p}, \Pi^{-1}(\Pi(\bar{a}_{i_p})))$ embeds in
$(\mf{A}_{i_r}, \Pi^{-1}(\Pi(\bar{a}_{i_r})))$, where $\Pi^{-1}$ is
the inverse permutation of $\Pi$. Then $(\mf{A}_{i_p}, \bar{a}_{i_p})$
embeds in $(\mf{A}_{i_r}, \bar{a}_{i_r})$ for each $p < r$, $p \ge 1$,
showing that $I$ is good. \epf

\begin{lemma}\label{lemma:sequence-with-all-permutations-is-good}
Let $J$ be an infinite sequence from $\cl{S}_k$ given by $J =
(\mf{A}_1, \Pi_1(\bar{a}_1)), \ldots,
(\mf{A}_1, \Pi_{k!}(\bar{a}_1)),$
$(\mf{A}_2, \Pi_1(\bar{a}_2)), \ldots,$
$(\mf{A}_2, \Pi_{k!}(\bar{a}_2)), \ldots$, where
$\mf{A}_i \in \cl{S}$, the components of $\bar{a}_i$ are distinct and
$\Pi_1, \ldots \Pi_{k!}$ are as in the previous lemma. Then $J$ is
good.
\end{lemma}
\emph{Proof}: Let $P_i$ be the set of the components of $\bar{a}_i$. 
Consider the sequence $(\mf{A}_i, P_i)_{i \ge 1}$ from
$\cl{S}^k$. Since $\cl{S}^k$ is a w.q.o. under $\hookrightarrow$,
there exists an infinite subsequence $(\mf{A}_{i, j}, P_{i, j})_{j \ge
1}$ s.t. $(\mf{A}_{i, p}, P_{i, p}) \hookrightarrow (\mf{A}_{i, r},
P_{i, r})$ for each $p < r$, $p \ge 1$. Then there must exist
permutations $\Pi_{j_{i, 1}}, \Pi_{j_{i, 2}}, \ldots$ s.t.
$(\mf{A}_{i, p}, \Pi_{j_{i, p}}(\bar{a}_{i, p})) \hookrightarrow
(\mf{A}_{i, r}, \Pi_{j_{i, r}}(\bar{a}_{i, r}))$ for each $p < r$,
$p \ge 1$. Then $J$ is a good sequence.\epf

We now complete the proof of
Lemma \ref{lemma:towards-two-equivalent-defns-of-P-wqo}. Assume that
we are given an infinite sequence $I_1$ from $\cl{S}_k$. Let
$c_1, \ldots, c_k$ be the constants of $\tau_k \setminus \tau$.  Let
$\Lambda$ be the finite set of all `types' of equalities/inequalities
between $c_1, \ldots, c_k$. Label each structure $(\mf{A},
a_1, \ldots, a_k)$ of $I_1$ with the (unique) type of $\Lambda$ that
is realized by the elements $a_1, \ldots, a_k$ in $\mf{A}$. Since
$\Lambda$ is finite, there must exist an infinite subsequence $I_2$ of
$I_1$ s.t.  the type of the equalities/inequalities between
$c_1, \ldots, c_k$ is the same in all the structures of
$I_2$. W.l.o.g. then, we can assume that the interpretations of
$c_1, \ldots, c_k$ are different from each other in all the structures
of $I_2$.  Then it follows from
Lemmas \ref{lemma:expansion-with-permutations}
and \ref{lemma:sequence-with-all-permutations-is-good} that $I_2$ must
be good. Whence $I_1$ is good.\epf

\section{Proof of Lemma \ref{lemma:composition-lemma-for-trees}}\label{appendix:tree-composition-lemma}

We present the proof
for (1); the proof for (2) is almost identical.

The base case is easy to check.  As the induction hypothesis, suppose
for $l < m$ that the duplicator has won in an $l$ round EF game
between $({\myr}_1, a_1, b_1)$ and $({\myr}_2, a_2, b_2)$ following
the strategy $\alpha$. For $i \in \{1, 2\}$, let $\mathsf{D}^i$ be the
set of elements chosen from ${\myr}_i$. Let
$g: \mathsf{D}^1 \rightarrow \mathsf{D}^2$ be the partial isomorphism
between $({\myr}_1, a_1, b_1)$ and $({\myr}_2, a_2, b_2)$. As the
induction step, suppose that at the end of the $(l+1)^{\text{th}}$
round, the elements chosen from ${\myr}_1$ and ${\myr}_2$ are
resp. $e_1$ and $e_2$. We will assume $e_i \notin \mathsf{D}^i$.  We
now show that $h = g \cup \{(e_1, e_2)\}$ is a partial isomorphism
between $({\myr}_1, a_1, b_1)$ and $({\myr}_2, a_2, b_2)$. There are
two cases here:

\underline{1. $e_1 \in \mathsf{U}_{s_1}$}: By definition of $\alpha$, it 
follows that $h$ restricted to $\mathsf{U}_{{\s}_1}$ is a partial
isomorphism between $({\s}_1, a_1)$ and $({\s}_2, a_2)$. Then the only
thing needed to be shown to complete the proof is that
${\myr}_1 \models (e_1 \leq c)$ iff ${\myr}_2 \models (h(e_1) \leq
h(c))$ for elements $c$ in
$\mathsf{D}^1 \cap \mathsf{U}_{{\f}_1}$. Observe that $h(c)$ must then
belong to $\mathsf{U}_{{\f}_2}$. Towards this, we see from the
construction of ${\myr}_1$ that (i) ${\myr}_1 \models (e_1 \leq c)$
iff ${\myr}_1 \models (e_1 \leq a_1)$ and (ii) ${\myr}_2 \models
(e_2 \leq h(c))$ iff ${\myr}_2 \models (e_2 \leq a_2)$. But since $h$
restricted to $\mathsf{U}_{{\s}_1}$ is a partial isomorphism between
$({\s}_1, a_1)$ and $({\s}_2, a_2)$, we have ${\myr}_1 \models
(e_1 \leq a_1)$ iff ${\myr}_2 \models (e_2 \leq a_2)$.

\underline{2. $e_1 \in \mathsf{U}_{{\f}_1}$}: By similar reasoning as above, 
we just need to show that ${\myr}_1 \models (c \leq e_1)$ iff
${\myr}_2 \models (h(c) \leq h(e_1))$ for elements $c$ in
$\mathsf{D}^1 \cap \mathsf{U}_{s_1}$. Observe that $h(c)$ must then
belong to $\mathsf{U}_{{\s}_2}$. By the construction of ${\myr}_1$, we
have (i) ${\myr}_1 \models (c \leq e_1)$ iff ${\myr}_1 \models (c \leq
a_1)$ and (ii) ${\myr}_2 \models (h(c) \leq e_2)$ iff
${\myr}_2 \models (h(c) \leq a_2)$. But since $h$ restricted to
$\mathsf{U}_{{\s}_1}$ is a partial isomorphism between $({\s}_1, a_1)$
and $({\s}_2, a_2)$, we have ${\myr}_1 \models (c \leq a_1)$ iff
${\myr}_2 \models (h(c) \leq a_2)$.

By induction then, $({\myr}_1, a_1, b_1) \equiv_m ({\myr}_2, a_2,
b_2)$. \epf

\section{Operations for constructing  new classes}\label{appendix:new-classes}

We give below the definitions of the disjoint union and cartesian
product operations.
\begin{enumerate}
\item The \emph{disjoint union} of $\mf{A}$ and $\mf{B}$, denoted
  $\mf{A} \sqcup \mf{B}$, is the structure $\mf{C}$ defined upto
  isomorphism as follows. Let $\mf{B}'$ be an isomorphic copy of
  $\mf{B}$ such that the universes of $\mf{A}$ and $\mf{B}'$ are
  disjoint. Then, (a) $\mathsf{U}_{\mf{C}} = \mathsf{U}_{\mf{A}} \cup
  \mathsf{U}_{\mf{B}'}$ (b) $\mf{C}(\mathsf{U}_{\mf{A}}) = \mf{A}$ and
  $M(\mathsf{U}_{\mf{B}'}) = \mf{B}'$ (c) for each predicate $R \in
  \tau$, of arity $k$, and for each $k$-tuple $\bar{a}$ of
  $\mathsf{U}_{\mf{C}}$ such that $\bar{a}$ contains at least one
  element from each of $\mathsf{U}_{\mf{A}}$ and
  $\mathsf{U}_{\mf{B}'}$, we have $\mf{C} \models \neg R(\bar{a})$.
\item The \emph{cartesian product} of $\mf{A}$ and $\mf{B}$, denoted
  $\mf{A} \times \mf{B}$, is the structure $\mf{C}$ defined as: (a)
  $\mathsf{U}_{\mf{C}} = \{ (a, b) \mid a \in \mathsf{U}_{\mf{A}}, b
  \in \mathsf{U}_{\mf{B}}\}$ (b) For each predicate $R \in \tau$, of
  arity $k$, for each $k$-tuple $\big((a_1, b_1), \ldots, (a_k,
  b_k)\big)$ of $\mathsf{U}_{\mf{C}}$, we have ${\mf{C}} \models
  R\big((a_1, b_1), \ldots, (a_k, b_k)\big)$ iff $\big( (a_1 = \cdots
  = a_k\, \wedge \, \mf{B} \models R(b_1, \ldots, b_k)) \bigvee
  (\mf{A} \models R(a_1, \ldots, a_k)\, \wedge \, b_1 = \cdots =
  b_k)\big)$.
\end{enumerate}

\section{Proof of Lemma \ref{lemma:baby-operations-preserve-pwqo-and-plogic}}
\label{appendix:baby-operations}

1) Suppose $\mc{P}_{wqo}(\cl{S}_i, k)$ holds for each $i \in \{1,
2\}$.  We will use the definition of $\mc{P}_{wqo}(\cl{S}, k)$ as
given by Lemma \ref{lemma:towards-two-equivalent-defns-of-P-wqo} for
reasoning about $!$ and $\sqcup$. For $\times$ and $\otimes$, we will
use the definition of $\mc{P}_{wqo}(\cl{S}, k)$ as given by
Definition \ref{defn:pwqo}.\\

Let $(\mf{A}_i, P_i)_{i \ge 1}$ be an infinite sequence from
$(!\cl{S}_1)^k$. Then consider the infinite sequence $(!\mf{A}_i,
P_i)_{i \ge 1}$ from $\cl{S}_1^k$. Since $\mc{P}_{wqo}(\cl{S}_1, k)$
holds, we have by
Lemma \ref{lemma:towards-two-equivalent-defns-of-P-wqo}, that
$\mc{P}_{wqo}(\cl{S}_1^k, 0)$ holds. Then there exist $i, j$ where $i
< j$ s.t. $(!\mf{A}_i, P_i) \hookrightarrow (!\mf{A}_j, P_j)$. Let $f$
be an embedding from $(!\mf{A}_i, P_i)$ to $(!\mf{A}_j, P_j)$. Then
from property P1 of $!$, it follows that $f$ is also an embedding from
$(\mf{A}_i, P_i)$ to $(\mf{A}_j, P_j)$.\\

Let $\cl{S} = \cl{S}_1 \sqcup \cl{S}_2$ and suppose $(\mf{A}_i,
P_i)_{i \ge 1}$ is an infinite sequence from $\cl{S}^k$. Let $\mf{A}_i
= \mf{B}_i^1 \sqcup \mf{B}_i^2$ where $\mf{B}_i^l \in \cl{S}_l$ for
$l \in \{1, 2\}$. Let $P_i^l$ be the subset of the universe of
$\mf{B}_i^l$ s.t. the disjoint union of $P_i^1$ and $P_i^2$ is $P_i$.
Construct the pair $H_i = \big( (\mf{B}_i^1, P_i^1), (\mf{B}_i^2,
P_i^2)\big)$ and consider the sequence $(H_i)_{i \ge 1}$. For each
$l \in \{1, 2\}$, since $\mc{P}_{wqo}(\cl{S}_l, k)$ holds, we have by
Lemma \ref{lemma:towards-two-equivalent-defns-of-P-wqo} that
$\mc{P}_{wqo}(\cl{S}_l^k, 0)$ holds. Then there must exist $i, j$
where $i < j$ s.t. $(\mf{B}_i^l, P_i^l) \hookrightarrow (\mf{B}_j^l,
P_j^l)$ for each $l \in \{1, 2\}$. Then from property P1 of $\sqcup$,
it follows that $(\mf{A}_i, P_i) \hookrightarrow (\mf{A}_j, P_j)$.\\

Let $\cl{S} = \cl{S}_1 \circledast \cl{S}_2$ for
$\circledast \in \{\times, \otimes\}$.  Suppose $(\mf{A}_i,
a_i^1, \ldots, a_i^k)_{i \ge 1}$ is an infinite sequence from
$\cl{S}_k$. Let $\mf{A}_i = \mf{B}_i^1 \circledast \mf{B}_i^2$ where
$\mf{B}_i^l \in \cl{S}_l$ for $l \in \{1, 2\}$. Let $b_i^{1, l} \ldots
b_i^{k, l}$ be the elements of the universe of $\mf{B}_i^l$ s.t.
$a_i^1 = (b_i^{1, 1}, b_i^{1, 2}), \ldots, a_i^k = (b_i^{k, 1},
b_i^{k, 2})$.  Constructing the pair $H_i = \big( (\mf{B}_i^1, b_i^{1,
1} \ldots b_i^{k, 1}), (\mf{B}_i^2, b_i^{1, 2} \ldots b_i^{k,
2})\big)$ and reasoning as in the previous paragraph, it follows that
there must exist $i, j$ where $i < j$ s.t. $(\mf{A}_i, a_i^1, \ldots,
a_i^k) \hookrightarrow (\mf{A}_j, a_j^1, \ldots, a_j^k)$.\\

2) The proofs for $\mc{P}_{logic}^{comp}$ can be done using similar
ideas as shown above. We just mention the computable functions in each
case. Let $\alpha_i$ be the computable function witnessing
$\mc{P}_{logic}^{comp}(\cl{S}_i, k)$ for $i \in \{1, 2\}$. Then
\begin{enumerate}
\item the function $\alpha_1$ witnesses $\mc{P}_{logic}^{comp}(\cl{S}_1, k)$.
\item the function $\alpha_1 + \alpha_2$ witnesses 
$\mc{P}_{logic}^{comp}(\cl{S}_1 \sqcup \cl{S}_2, k)$.
\item the function $\alpha_1 \times \alpha_2$ witnesses 
$\mc{P}_{logic}^{comp}(\cl{S}_1 \circledast \cl{S}_2, k)$ for
$\circledast \in \{\times, \otimes\}$.
\end{enumerate}

\end{appendix}

\end{document}